\documentclass[12pt,a4paper]{article}
\pdfoutput=1
\usepackage{jheppub,amsmath,amssymb,color,youngtab}
\usepackage[nottoc]{tocbibind}
\usepackage{natbib}
\newcommand{\bea}{\begin{eqnarray}}
\newcommand{\eea}{\end{eqnarray}}

\newcommand{\rf}[2]{\left(#1\right)_{#2}}

\begin{document}
\title{Bulk Reconstruction in Bilocal Holography}
\author[a,b]{Robert de Mello Koch,}
\author[a]{Animik Ghosh,}
\author[a,c]{Minkyoo Kim}
\author[a,b]{ and Anik Rudra,}
\affiliation[a]{School of Science, Huzhou Normal University, Huzhou 313000, China}
\affiliation[b]{Mandelstam Institute for Theoretical Physics, School of Physics, University of the Witwatersrand, Private Bag 3, Wits 2050, South Africa}
\affiliation[c]{Department of Physics and Astronomy \& Center for Theoretical Physics, Seoul National University, Seoul 08826, Korea}
\emailAdd{robert@zjhu.edu.cn,animikghosh@gmail.com,minkyookim@snu.ac.kr,\\anikrudra23@gmail.com}
\date{April 2026}
\abstract{Bilocal holography provides a constructive approach to the higher-spin gravity theories dual to vector-model conformal field theories. Its central advantage is that it is completely gauge fixed and formulated entirely in terms of physical degrees of freedom. We derive a remarkably local bulk reconstruction formula and demonstrate its agreement with standard bulk reconstruction, after the same boundary data and gauge-fixed variables have been identified. We further clarify how subregion duality is realized in this framework.}
\maketitle

\section{Introduction}

The AdS/CFT correspondence asserts an equivalence between a higher-dimensional quantum gravity theory and a lower-dimensional non-gravitational quantum field theory \cite{Maldacena:1997re,Gubser:1998bc,Witten:1998qj}. This is striking because quantum field theory is comparatively well understood, while quantum gravity remains opaque. AdS/CFT therefore offers a route to the microscopic structure of quantum gravity. Yet, despite the substantial evidence for the correspondence, we have no complete understanding of how the boundary degrees of freedom reorganize into a bulk gravitational description. In this paper we approach this question using collective field theory \cite{Jevicki:1979mb,Jevicki:1980zg}, a framework proposed as a constructive route to holography in~\cite{Das:2003vw,Das:1990kaa}.

Part of the challenge is to identify variables in which bulk emergence is explicit rather than inferred indirectly from symmetry, correlators, or consistency conditions. Bilocal holography \cite{Das:2003vw} is especially attractive from this point of view. It is a constructive approach to the duality \cite{Klebanov:2002ja,Sezgin:2002rt} between vector-model conformal field theories and higher-spin gravity \cite{Vasiliev:1990en,Vasiliev:2003ev}. Rather than beginning from bulk fields and asking how they are encoded on the boundary, bilocal holography starts with the gauge-invariant degrees of freedom of the boundary theory itself, reorganized as collective fields, and asks whether the bulk spacetime, its local fields, and its redundancy can be derived directly from that reorganization. In this sense, bilocal holography is not merely a check of the duality; it is a proposal for how the duality is built.

What makes this proposal particularly compelling is that it works at the level of \emph{independent dynamical degrees of freedom}. In the vector-model example, once the conserved currents of the conformal field theory are reduced to their physical components, one finds precisely the same number of degrees of freedom as in the completely gauge-fixed higher-spin theory in AdS~\cite{Metsaev:1999ui,Metsaev:1999gz,Metsaev:1999gw,Metsaev:1999kb,Metsaev:2005ws}. The bilocal collective field then furnishes an explicit intertwiner between these two descriptions (see also~\cite{Czech:2015qta,Czech:2016xec}). This is a highly nontrivial statement. One is not merely matching symmetries or correlators abstractly, but identifying the physical variables on the two sides of the correspondence. In this form the holographic map is concrete enough to support explicit bulk reconstruction, to realize the expected localization of information, and to make the kinematics behind entanglement-wedge reconstruction unusually explicit.

Bilocal holography is a concrete realization of the broader collective-field program \cite{Jevicki:1979mb,Jevicki:1980zg}. It provides explicit formulae for higher-spin gravity fields in terms of operators of the conformal field theory, thereby realizing a construction of precursors---that is, bulk fields represented directly in terms of boundary operators \cite{Polchinski:1999yd}. One key technical input that makes this possible is the remarkably detailed light-cone analysis of higher-spin gravity due to Metsaev \cite{Metsaev:1999ui,Metsaev:1999gz,Metsaev:1999gw,Metsaev:1999kb,Metsaev:2005ws}, which achieves a complete gauge fixing to physical degrees of freedom and allows a direct comparison with the reduced degrees of freedom of the vector model. The original map was constructed in \cite{deMelloKoch:2010wdf}, where it was shown that the full set of conformal generators is mapped into the corresponding higher-spin generators. Since then, bilocal holography has been developed in a number of directions \cite{Jevicki:2011ss,Jevicki:2011aa,deMelloKoch:2012vc,deMelloKoch:2014mos,deMelloKoch:2014vnt,Mulokwe:2018czu,deMelloKoch:2018ivk,deMelloKoch:2021cni,deMelloKoch:2022sul,Johnson:2022cbe,deMelloKoch:2023ngh,deMelloKoch:2023ylr,Mintun:2014gua,Aharony:2020omh}.

Conceptually, bilocal holography has two essential ingredients. First, there is a change of field variables, from the original microscopic fields to collective bilocals. This reorganizes the perturbative expansion so that the loop parameter changes from \(\hbar\) to \(1/N\). Second, there is a change of spacetime coordinates, which solves the Clebsch--Gordan problem of passing from the tensor-product basis natural for the bilocal field, to the direct-sum basis natural for the higher-spin bulk fields. The second step is highly nontrivial: it is this change of variables that converts the kinematics of a bilocal object into the kinematics of a local bulk field with an additional spin coordinate. In the conformal field theory one performs a reduction to independent degrees of freedom by eliminating components of the conserved current related by the conservation equation. The reduction is achieved by employing equal-$x^+$ bilocal collective fields~\cite{deMelloKoch:2023ngh}. The conformal generators are those of the reduced theory. In the higher spin gravity, one performs a complete gauge fixing in lightcone gauge. The generators of the AdS isometries do not preserve the gauge choice, so that they must be composed with compensating gauge transformations that restore the gauge. These conformal generators of the higher spin gravity were completely evaluated in the beautiful paper~\cite{Metsaev:1999ui}. The paper~\cite{deMelloKoch:2010wdf} established a relationship between the bilocal coordinates and those of the AdS bulk, under which the complete set of conformal generators is mapped into the generators of the higher-spin theory, for all spins. 

In this paper we focus on the duality between the free $O(N)$ vector model CFT$_3$ and the minimal higher-spin gravity on AdS$_4$. The bulk theory contains a scalar together with gauge fields of every even integer spin, while the single-trace primaries of the boundary theory consist of a scalar of dimension $\Delta=1$ and a tower of conserved currents of spin $s$ and dimension $s+1$, again for every even integer $s$. After gauge fixing and reduction to independent degrees of freedom, the complete set of conformal field theory degrees of freedom is packaged into a single $O(N)$-invariant equal-time bilocal field, while on the gravity side one is left with two physical polarizations of each even-spin gauge field, together with the bulk scalar. Since the bulk analysis is naturally performed in light-cone gauge, it is natural to quantize both theories on the light front, taking $x^+$ to play the role of time.

In terms of the fundamental scalar $\phi^a$ of the $O(N)$ model, the equal-time bilocal field is
\begin{align}
\sigma(x^+,x_1^-,x_1,x_2^-,x_2)=\phi^a(x^+,x_1^-,x_1)\,\phi^a(x^+,x_2^-,x_2)\, 
\end{align}
Here $x^\pm$ are light-cone coordinates on the CFT side, and $x$ is the single transverse coordinate. The bilocal field develops a large-$N$ expectation value $\sigma_0$, and expanding about this background defines the fluctuation field $\eta$,
\begin{align}
\sigma(x^+,x_1^-,x_1,x_2^-,x_2)=\sigma_0(x^+,x_1^-,x_1,x_2^-,x_2)+
\eta(x^+,x_1^-,x_1,x_2^-,x_2)\, .
\end{align}
It is the fluctuation $\eta$, depending on five coordinates, that is identified with the dynamical degrees of freedom of the dual higher-spin theory.

On the bulk side it is convenient to package the scalar and the full tower of higher-spin fields into a single master field by introducing an extra auxiliary angular coordinate \(\theta\),
\begin{align}
\Phi(X^+,X^-,X,Z,\theta)&=\sum_{s=0}^{\infty}\left[
\frac{A^{XX\cdots XX}(X^+,X^-,X,Z)}{Z}\cos(2s\theta)\right.\nonumber\\
&\left.+\frac{A^{XX\cdots XZ}(X^+,X^-,X,Z)}{Z}\sin(2s\theta)\right].
\end{align}
The fields appearing here are tangent-space fields; see \cite{Metsaev:1999ui} for a useful discussion. As emphasized in \cite{deMelloKoch:2021cni}, the map between the bilocal fluctuation and the bulk master field is most naturally written in a mixed position/momentum representation. Since both theories are invariant under translations in \(x^-\) and \(X^-\), one Fourier transforms in these directions, trading \((x_1^-,x_2^-)\) for \((p_1^+,p_2^+)\) and \(X^-\) for \(P^+\). The bilocal reconstruction of the bulk master field then takes the form
\begin{equation}
\Phi^{\rm biloc}(X^+,P^+,X,Z,\theta)=4\pi\sqrt{p_1^+p_2^+}\, \eta(x^+,p_1^+,x_1,p_2^+,x_2),\label{eq:Phi-eta-map}
\end{equation}
together with the coordinate transformation
\begin{align}
X^+&=x^+,\qquad P^+=p_1^+ + p_2^+,\qquad
X=\frac{p_1^+x_1+p_2^+x_2}{p_1^+ + p_2^+},\nonumber\\[2mm]
Z&=\frac{\sqrt{p_1^+p_2^+}}{p_1^+ + p_2^+}|x_1-x_2|,\qquad
\theta=2\arctan\sqrt{\frac{p_2^+}{p_1^+}}.\label{coordtrans}
\end{align}
This is the basic bilocal holography map. The coordinate transformation is easily inverted to find
\begin{align}
x^+&=X^+,\qquad p_1^+=P^+ \cos^2\frac{\theta}{2}\qquad p_2^+=P^+\sin^2\frac{\theta}{2},\nonumber\\[2mm]
x_1&=X+Z\tan\frac{\theta}{2},\qquad
x_2=X-Z\cot\frac{\theta}{2}.\label{invcoordtrans}
\end{align}
where we assign the labels 1 and 2 such that $x_1\ge x_2$.

Once this map is in hand, a number of stringent tests become possible. One particularly sharp question is whether bilocal holography realizes the expected localization of information in quantum gravity. The holography of information~\cite{Laddha:2020kvp,Chowdhury:2020hse,Raju:2020smc,Raju:2021lwh,SuvratYouTube} asserts that information in a theory of quantum gravity is localized in a manner radically different from that in ordinary quantum field theory or classical field theory. In particular, complete knowledge of the state in an arbitrarily small neighbourhood of the boundary, on a given Cauchy slice, is sufficient to determine the state on the entire slice. Using the ordinary operator product expansion, Ref.~\cite{deMelloKoch:2022sul} showed that bilocal holography provides a concrete realization of this principle, with the expected localization properties reproduced in detail.

The value of this framework is therefore not merely that the duality works, but that the mechanism by which it works becomes visible. Bilocal holography provides a setting in which one can follow explicitly how the independent gauge-invariant degrees of freedom of the boundary theory reorganize themselves into local bulk variables. This constructive perspective makes it especially well suited to questions of bulk reconstruction and subregion duality, which are the central themes of this paper.

We begin in the next section by deriving the bilocal bulk reconstruction formula. The derivation proceeds by expanding the bilocal in terms of single-trace primaries and their descendants, and then resumming the descendant contributions. This gives a reconstruction formula for the bulk master field directly in terms of the boundary primaries. The resulting expression is remarkably local. In Section~\ref{sec:bulk-equation}, reviewing and sharpening results of Refs.~\cite{deMelloKoch:2010wdf,deMelloKoch:2014vnt,Mintun:2014gua}, we show that the collective field $\Phi^{\rm biloc}$ obeys the correct bulk equations of motion and the correct boundary conditions. By uniqueness, it must therefore agree with bulk fields reconstructed by other methods once the same reduced variables, boundary data and Lorentzian prescription are used. In Section~\ref{HKLLequivalence} we verify this explicitly by comparing with HKLL reconstruction~\cite{Hamilton:2006az,Heemskerk:2012mn}; the agreement is exact for the reduced master field. Finally, in Section~\ref{sec:subregion-modular} we turn to subregion duality. We show that restricting the bilocal to a single boundary subregion reconstructs precisely the corresponding entanglement wedge~\cite{deMelloKoch:2021cni}, in accord with general expectations from entanglement wedge reconstruction~\cite{Czech:2012bh,Headrick:2014cta,Wall:2012uf,Jafferis:2015del,Dong:2016eik,Cotler:2017erl}. We conclude with a discussion of the implications of our study in Section \ref{conclusions} and collect technical details in the appendices.

\section{Bilocal Bulk Reconstruction}\label{BBR}

In this section our goal is to write down the formula reconstructing the bulk master field $\Phi(X^+,X^-,X,Z,\theta)$ in terms of the bilocal fluctuation $\eta(x^+,p_1^+,x_1,p_2^+,x_2)$. The field mapping of bilocal holography \eqref{eq:Phi-eta-map} is formulated in the mixed representation, in which $x^-$ has been Fourier transformed to $p^+$, as follows
\begin{equation}
\eta(x^+,p_1^+,x_1,p_2^+,x_2)=\int dx_1^-\,dx_2^-\,e^{i p_1^+x_1^-+i p_2^+x_2^-}\,\eta(x^+,x_1^-,x_1,x_2^-,x_2).
\label{eq:auto:0002}
\end{equation}
In subsection \ref{etaprimary} we use the OPE to express this fluctuation in terms of the single trace primary operators and their descendants. This result is then used in subsection \ref{bilocalbulk} to derive the explicit bulk reconstruction formulas of bilocal holography.

\subsection{Bilocal fluctuation in terms of primaries}\label{etaprimary}
The conformal field theory we study is the free $O(N)$ vector model. For this theory, the equal-$x^+$ OPE is (see Appendix A of \cite{deMelloKoch:2022sul})
\begin{equation}
\sum_{a=1}^N:\phi^a(x+y)\phi^a(x-y):=\sum_{s=0}^{\infty}\sum_{d=0}^{\infty}c_{sd}\,\bigl(y^\mu \partial_\mu \bigr)^{2d}\,y^{\mu_1}\cdots y^{\mu_{2s}}\,j^{(2s)}_{\mu_1\cdots\mu_{2s}}(x),\label{eq:full-OPE}
\end{equation}
where
\begin{equation}
c_{0d}=\frac{1}{2^{2d} (d!)^2} \qquad {\rm and}\qquad
c_{sd}=\frac{(2 s)! (4 s-1)!!}{d! 2^{2 d+4 s-1} (d+2 s)!}\qquad\qquad s>0\label{eq:csd}
\end{equation}
In the OPE we use the coordinates
\begin{equation}
x^\mu=\frac{x_1^\mu+x_2^\mu}{2},\qquad\qquad y^\mu=\frac{x_1^\mu-x_2^\mu}{2}.
\label{eq:auto:0003}
\end{equation}
On the equal-$x^+$ slice, we have $y^+=0$. Thus, only $x$ and $-$ polarizations appear on the right hand side of \eqref{eq:full-OPE}. As the current is totally symmetric, all that matters is how many indices are $-$ and how many are $x$. This motivates the useful notation
\begin{equation}
j_{2s}^{(m)}(x^+,x^-,x)\equiv j_{\underbrace{-\cdots -}_{m}\underbrace{x\cdots x}_{2s-m}}(x^+,x^-,x),\qquad m=0,1,\dots,2s.\label{eq:auto:0004}
\end{equation}
In terms of this notation we have
\begin{equation}
y^{\mu_1}\cdots y^{\mu_{2s}}j_{\mu_1\cdots\mu_{2s}}(x)=\sum_{m=0}^{2s}\binom{2s}{m}(y^-)^m y^{\,2s-m}\,
j_{2s}^{(m)}(x^+,x^-,x),\label{eq:tensor-decomposition}
\end{equation}
where we used the fact that the number of ways to choose $m$ $-$ indices from the total number of $2s$ indices is given by $\binom{2s}{m}$. Applying the equal-$x^+$ OPE to fluctuation $\eta$ we obtain
\begin{equation}
\eta_{(2s)}(x^+,x^-,y^-,x,y)=\sum_{d=0}^\infty c_{sd}\,\bigl(y^-\partial_{x^-}+y\partial_x\bigr)^{2d}\sum_{m=0}^{2s}
\binom{2s}{m}\,(y^-)^m y^{\,2s-m}\,j_{2s}^{(m)}(x^+,x^-,x).\label{eq:position-space-OPE}
\end{equation}
The subscript $(2s)$ indicates that we are working with a single term from the sum over $s$ on the right hand side of \eqref{eq:full-OPE}. The complete fluctuation $\eta$ is obtained by summing over $s$. This formula expresses the position space fluctuation $\eta$ in terms of the single trace primaries and their descendants. The bilocal holography mapping between the bilocal fluctuation and the bulk master field \eqref{eq:Phi-eta-map} uses fields in a mixed position/momentum space representation. This motivates the Fourier transform
\begin{equation}
\eta_{(2s)}(x^+,P^+,q^+;x,y)=2\int dx^-\,dy^-\,e^{iP^+x^-+iq^+y^-}\,\eta_{(2s)}(x^+,x^-,y^-,x,y).\label{eq:mixed-eta}
\end{equation}
Here we have chosen to use the variables $P^+=p_1^++p_2^+$ and $q=p_1^+-p_2^+$. This choice will prove convenient when we express the reconstruction formula in terms of bulk variables. Plugging \eqref{eq:position-space-OPE} into \eqref{eq:mixed-eta} we obtain
\begin{eqnarray}
\eta_{(2s)}(x^+,P^+,q^+;x,y)&=&2\int dx^-\,dy^-\,e^{iP^+x^-+iq^+y^-}\,\sum_{d=0}^\infty c_{sd}\,\bigl(y^-\partial_{x^-}+y\partial_x\bigr)^{2d}\cr\cr
&&\quad\times\,\sum_{m=0}^{2s}\binom{2s}{m}\,(y^-)^m y^{\,2s-m}\,j_{2s}^{(m)}(x^+,x^-,x)\label{currinprog}
\end{eqnarray}
We will now evaluate the sum over descendants. This sum over $d$ is a sum of powers of derivatives. Derivatives with respect to $x^-$ are easily evaluated by integrating by parts and allowing them to act on the exponential. To evaluate the derivatives with respect to $x$ it is useful to use the Fourier transform of the current
\begin{equation}
j_{2s}^{(m)}(x^+,x^-,x)=\int \frac{dP^{\prime +}\,dk}{(2\pi)^2}\,e^{-iP^{\prime +}x^- -ikx}\,j_{2s}^{(m)}(x^+,P^{\prime +},k).\label{eq:j-inverse}
\end{equation}
We then obtain
\begin{align}
\eta_{(2s)}(x^+,P^+,q^+;x,y)&=2\sum_{d=0}^{\infty}c_{sd}\sum_{m=0}^{2s}\binom{2s}{m}y^{\,2s-m}
\int dx^-\,dy^-\,e^{iP^+x^-+iq^+y^-}(y^-)^m\nonumber\\
&\times\int \frac{dP^{\prime +}\,dk}{(2\pi)^2}\,e^{-iP^{\prime +}x^- -ikx}\,\bigl[-i(P^{\prime +}y^-+ky)\bigr]^{2d}j_{2s}^{(m)}(x^+,P^{\prime +},k).\label{eq:after-diagonalization}
\end{align}
The $x^-$ integral produces a delta function in $P^{\prime +}$ making the  integral over $P^{\prime +}$ completely straightforward. The result is
\begin{align}
\eta_{(2s)}(x^+,P^+,q^+;x,y)&=2\sum_{m=0}^{2s}\binom{2s}{m}y^{\,2s-m}\int \frac{dk}{2\pi}\,e^{-ikx}\,
j_{2s}^{(m)}(x^+,P^+,k)\nonumber\\
&\qquad\times\int dy^-\,e^{iq^+y^-}(y^-)^m\sum_{d=0}^{\infty}c_{sd}\,\bigl[-i(P^+y^-+ky)\bigr]^{2d}.
\label{eq:before-Ism}
\end{align}
Introduce the quantities $\beta\equiv ky$ and $\nu\equiv \frac{q^+}{P^+}$. We can now write
\begin{align}
\eta_{(2s)}(x^+,P^+,q^+;x,y)&=2\sum_{m=0}^{2s}\binom{2s}{m}y^{\,2s-m}\int \frac{dk}{2\pi}\,e^{-ikx}\,\nonumber\\
&\times j_{2s}^{(m)}(x^+,P^+,k)\,\mathcal I_{s,m}\!\left(\nu,\beta;P^+\right).\label{eq:mixed-master-derived}
\end{align}
where we have introduced the function
\begin{equation}
\mathcal I_{s,m}(\nu,\beta;P^+)\equiv\int dy^-\,e^{i\nu P^+y^-}(y^-)^m\sum_{d=0}^{\infty}c_{sd}\,
\bigl[-i(P^+y^-+\beta)\bigr]^{2d}.\label{eq:Ism-definition-again}
\end{equation}
To perform the sum over descendants in \eqref{eq:Ism-definition-again} introduce the functions
\begin{equation}
F_0(u)\equiv \sum_{d=0}^\infty c_{0d}\,u^{2d}=\sum_{d=0}^\infty\,\frac{1}{2^{2d} (d!)^2}\,u^{2d},\label{eq:def-F0}
\end{equation}
\begin{equation}
F_s(u)\equiv \sum_{d=0}^\infty c_{sd}\,u^{2d}=\sum_{d=0}^\infty \frac{(2 s)! (4 s-1)!!}{d! 2^{2 d+4 s-1} (d+2 s)!}\,u^{2d}\qquad\qquad s>0.\label{eq:def-Fs}
\end{equation}
In terms of these functions we have
\begin{equation}
\mathcal I_{s,m}(\nu,\beta;P^+)=\int dy^-\,e^{i\nu P^+y^-}(y^-)^m F_s\bigl(-i(P^+y^-+\beta)\bigr)\qquad
s>0.\label{eq:Ism-Fs}
\end{equation}
Noting that
\begin{equation}
I_{2s}(u)=\sum_{d=0}^\infty\frac{1}{d!\,\Gamma(d+2s+1)}\left(\frac{u}{2}\right)^{2d+2s}=\sum_{d=0}^\infty\frac{1}{d!\,(d+2s)!}\left(\frac{u}{2}\right)^{2d+2s},
\label{eq:auto:0005}
\end{equation}
we obtain
\begin{eqnarray}
F_0(u)=I_0(u)\qquad&\Rightarrow&\qquad F_0(-i\xi)=J_0(\xi)\cr\cr
F_s(u)=\frac{(2s)!(4s-1)!!}{2^{2s-1}}\,u^{-2s}I_{2s}(u)\qquad&\Rightarrow&\qquad
F_s(-i\xi)=\frac{(2s)!(4s-1)!!}{2^{2s-1}}\,\frac{J_{2s}(\xi)}{\xi^{2s}},\nonumber
\end{eqnarray}
Change variables from $y^-$ to $\xi \equiv P^+y^-+ky$ and use
\begin{equation}
y^-=\frac{\xi-\beta}{P^+},\qquad dy^-=\frac{d\xi}{P^+},\qquad e^{iq^+y^-}=e^{i\nu (\xi-\beta)}=
e^{-i\nu \beta}e^{i\nu\xi},\label{eq:change-vars}
\end{equation}
to obtain
\begin{align}
\mathcal I_{s,m}(\nu,ky;P^+)&=\frac{e^{-i\nu ky}}{(P^+)^{m+1}}\int d\xi\,e^{i\nu\xi}\,(\xi-ky)^m\,F_s(-i\xi)\qquad
s>0.\label{eq:Ism-after-xi}
\end{align}
We can evaluate the Fourier transform as follows\footnote{This Fourier-transform formula assumes $|\nu|<1$; outside this interval it vanishes. For $s\ge 1$, the kernel $J_{2s}(\xi)/\xi^{2s}$ is locally regular at $\xi=0$ and decays as $|\xi|^{-2s-\frac12}$ at infinity, so its Fourier transform may be interpreted as an ordinary improper integral. In the scalar case $s=0$, the transform is naturally understood in the oscillatory/distributional sense.}
\begin{eqnarray}
&&\int d\xi\,e^{i\nu\xi}\,(\xi-ky)^m\,F_s(-i\xi)\,\,=\,\,
\frac{(2s)!(4s-1)!!}{2^{2s-1}}\int d\xi\,e^{i\nu\xi}\,(\xi-ky)^m\,\,\frac{J_{2s}(\xi)}{\xi^{2s}}\cr\cr
&=&\frac{(2s)!(4s-1)!!}{2^{2s-1}}\left(-i\frac{\partial}{\partial \nu}-ky\right)^m\left[
\frac{2^{1-2 s}\sqrt{\pi} \left(1-\nu ^2\right)^{2 s-\frac{1}{2}}}{\Gamma \left(2 s+\frac{1}{2}\right)}\right]
\qquad s>0\label{eq:auto:0065}
\end{eqnarray}
which then implies that
\begin{align}
\mathcal I_{s,m}(\nu,ky;P^+)&=\frac{e^{-i\nu ky}(2s)!(4s-1)!!}{2^{2s-1}(P^+)^{m+1}}\nonumber\\
&\times\left(-i\frac{\partial}{\partial \nu}-ky\right)^m\left[
\frac{2^{1-2 s}\sqrt{\pi} \left(1-\nu ^2\right)^{2 s-\frac{1}{2}}}{\Gamma \left(2 s+\frac{1}{2}\right)}\right]\qquad s>0.
\label{eq:auto:0060}
\end{align}
Performing the same steps in the scalar $s=0$ case gives
\begin{equation}
\mathcal I_{0,0}(\nu,ky;P^+)\,\,=\,\, \frac{2e^{-i\nu ky}}{P^+\sqrt{1-\nu^2}}\,.\label{eq:auto:0007}
\end{equation}
We have assumed that $\nu\in(-1,1)$. This is justified in the physical kinematic regime $p_1^+,p_2^+>0$ since
\begin{equation}
\nu=\frac{p_1^+-p_2^+}{p_1^++p_2^+}.
\label{eq:auto:0008}
\end{equation}
In the formula below $j_{0}^{(0)}(x^+,x^-,x)$ stands for the scalar primary of dimension 1
\begin{equation}
j_{0}^{(0)}(x^+,x^-,x)\,\,=\,\,:\phi^a\phi^a(x^+,x^-,x):
\label{eq:auto:0009}
\end{equation}
Thus, we finally obtain
\begin{eqnarray}
\eta_{(2s)}(x^+,P^+,q^+;x,y)&=&2\sum_{m=0}^{2s}\frac{2^{\,2-2s}(2s)!}{(P^+)^{m+1}}\binom{2s}{m}y^{\,2s-m}\int \frac{dk}{2\pi}\,e^{-ikx}\,j_{2s}^{(m)}(x^+,P^+,k)\cr\cr
&&\times e^{-i\nu ky}\left(-i\frac{\partial}{\partial \nu}-\beta\right)^m
\left(1-\nu ^2\right)^{2 s-\frac{1}{2}}\qquad s>0\cr\cr
\eta_{(0)}(x^+,P^+,q^+;x,y)&=&2\int \frac{dk}{2\pi}\,e^{-ikx}\,
j_{0}^{(0)}(x^+,P^+,k)\,\frac{2e^{-i\nu ky}}{P^+\sqrt{1-\nu^2}}.
\label{eq:eta-mixed-final}
\end{eqnarray}
These formulas can be simplified further. Since $\beta$ is independent of $\nu$
\begin{equation}
e^{-i\nu\beta}\left(-i\frac{\partial}{\partial \nu}-\beta\right)=\left(-i\frac{\partial}{\partial \nu}\right)e^{-i\nu\beta},
\label{eq:auto:0012}
\end{equation}
and hence, by iteration,
\begin{equation}
e^{-i\nu\beta}\left(-i\frac{\partial}{\partial \nu}-\beta\right)^m f(\nu)=\left(-i\frac{\partial}{\partial \nu}\right)^m
\Bigl[e^{-i\nu\beta}f(\nu)\Bigr].\label{eq:conjugation}
\end{equation}
Next, let $a=2s-\frac12$. The standard Jacobi-polynomial identity gives
\begin{equation}
\frac{d^r}{d\nu^r}(1-\nu^2)^a=(-2)^r r!\,(1-\nu^2)^{\,a-r}P_r^{(a-r,a-r)}(\nu).\label{eq:Jacobi-derivative}
\end{equation}
Using the product rule together with
\begin{equation}
\left(-i\frac{\partial}{\partial \nu}\right)^{m-r}e^{-i\nu\beta}=(-\beta)^{m-r}e^{-i\nu\beta},
\label{eq:auto:0013}
\end{equation}
we obtain
\begin{align}
\left(-i\frac{\partial}{\partial \nu}\right)^m\Bigl[e^{-i\nu\beta}(1-\nu^2)^a\Bigr]&=e^{-i\nu\beta}
\sum_{r=0}^m \binom{m}{r} (-\beta)^{m-r} (-i)^r\frac{d^r}{d\nu^r}(1-\nu^2)^a\nonumber\\[4pt]
&=e^{-i\nu\beta}\sum_{r=0}^m\binom{m}{r}(-\beta)^{m-r}(2i)^r r!\,(1-\nu^2)^{\,a-r} P_r^{(a-r,a-r)}(\nu).
\label{eq:Jacobi-sum-derivative}
\end{align}
Using this last formula in \eqref{eq:eta-mixed-final}, we obtain
\begin{eqnarray}
\eta_{(2s)}(x^+,P^+,q^+;x,y)&=&2^{\,3-2s}(2s)!\sum_{m=0}^{2s}\sum_{r=0}^m\binom{2s}{m}\binom{m}{r}
\frac{r! (2i)^r(-i)^{\,m-r}y^{\,2s-r}}{(P^+)^{m+1}}\cr\cr
&&\!\!\!\!\!\!\!\times(1-\nu^2)^{\,2s-r-\frac12}P_r^{\left(2s-r-\frac12,\;2s-r-\frac12\right)}(\nu)\,
\partial_x^{\,m-r}j^{(m)}_{2s}(x^+,P^+;x+\nu y),\cr\cr
\eta_{(0)}(x^+,P^+,q^+;x,y)&=&\frac{2^2}{P^+\sqrt{1-\nu^2}}\, j_0^{(0)}(x^+,P^+;x+\nu y).
\label{eq:eta-xspace-final}
\end{eqnarray}
Thus, the bilocal fluctuation in the mixed position/momentum space representation, can be expressed in terms of the single trace primaries and their descendants as
\begin{equation}
\eta(x^+,p_1^+,x_1,p_2^+,x_2)=\sum_{s=0}^\infty \eta_{(2s)}(x^+,p_1^++p_2^+,p_1^+-p_2^+;\frac{x_1+x_2}{2},\frac{x_1-x_2}{2})
\label{eq:auto:0014}
\end{equation}
with $\eta_{(2s)}(x^+,P^+,q^+;x,y)$ given above.

\subsection{The bilocal-to-bulk map}\label{bilocalbulk}
Recall the relation \eqref{eq:Phi-eta-map} between the bilocal reconstruction of the bulk master field and the collective field fluctuation 
\begin{equation}
\Phi^{\rm biloc}(X^+,P^+,X,Z,\theta)=2\pi P^+\sin\theta\,\eta(x^+,p_1^+,x_1,p_2^+,x_2),
\end{equation}
The bulk master field packages bulk fields of different spins
\begin{equation}
\Phi(X^+,P^+,X,Z,\theta)=\sum_{r=0}^\infty\left[\cos(2r\theta)\,\frac{\Phi_{2r}^{(c)}(X^+,P^+,X,Z)}{Z}+
\sin(2r\theta)\,\frac{\Phi_{2r}^{(s)}(X^+,P^+,X,Z)}{Z}\right].\nonumber
\end{equation}
Making the change of field basis $\Phi_{2r}^{(c)}=\Phi_{2r}^{(+)}+\Phi_{2r}^{(-)}$ $\Phi_{2r}^{(s)}=i\Phi_{2r}^{(+)}-i\Phi_{2r}^{(-)}$ we obtain
\begin{equation}
\Phi(X^+,P^+,X,Z,\theta)=\sum_{s=0}^\infty\left[e^{2is\theta}\,\frac{\Phi_{2s}^{(+)}(X^+,P^+,X,Z)}{Z}+
e^{-2is\theta}\,\frac{\Phi_{2s}^{(-)}(X^+,P^+,X,Z)}{Z}\right].\nonumber
\end{equation}
This makes it clear that the spin-$2s$ helicity components are obtained by projection
\begin{align}
\Phi_{2s}^{(+)}(X^+,P^+,X,Z)&=\frac{Z}{2\pi}\int_0^{2\pi} d \theta\; e ^{-2 i  s\theta}\Phi(X^+,P^+,X,Z,\theta),
\nonumber\\
\Phi_{2s}^{(-)}(X^+,P^+,X,Z)&=\frac{Z}{2\pi}\int_0^{2\pi} d \theta\; e ^{+2 i  s\theta}\Phi(X^+,P^+,X,Z,\theta).
\label{eq:projection-formula}
\end{align}
We will now derive the formula for the bilocal bulk reconstruction. Using the formula for $\eta$ obtained in the previous subsection, the reconstructed master field becomes
\begin{align}
\Phi^{\rm biloc}\big|_{\eta^{(2s)}}&=2\pi\,2^{3-2s}(2s)!\sum_{m=0}^{2s}\sum_{r=0}^{m}\binom{2s}{m}\binom{m}{r}
(2 i )^r(- i )^{m-r}r!\frac{Z^{2s-r}}{(P^+)^{m}}\nonumber\\
&\hspace{3cm}\times (\sin\theta)^{2s-r} P_r^{\left(2s-r-\frac12,\,2s-r-\frac12\right)}(\cos\theta)
\,\partial_X^{m-r}j_{2s}^{(m)}(x^+,P^+;X).\label{eq:master-from-current-sector}
\end{align}
To obtain this formula we have used 
\begin{equation}
\nu=\cos\theta,\qquad y=Z/\sin\theta,\qquad x+\nu y=X, \qquad 1-\nu^2=\sin^2\theta
\end{equation} 
which follow immediately from the bilocal coordinate mapping \eqref{coordtrans} and \eqref{invcoordtrans}. Notice that $\Phi^{\rm biloc}\big|_{\eta^{(2s)}}$ is the contribution to the reconstructed bulk field from $\eta^{(2s)}$. To obtain the full reconstruction we need to sum $\Phi^{\rm biloc}\big|_{\eta^{(2s)}}$ over $s$. We now project out the spin $2s$ harmonic from $\Phi^{\rm biloc}|_{\eta_{(2s)}}$. In general, there is a non-zero result when projecting the spin-$2s$ Fourier harmonic out of $\Phi^{\rm biloc}|_{\eta_{(2s')}}$ with $s\neq s'$, provided $s<s'$. This follows because, as will become clear below, the spin-$2s'$ block has finite angular Fourier support
\begin{equation}
e^{2in\theta},\qquad n=-s',-s'+1,\ldots,s'-1,s'.
\end{equation}
Thus, by projecting out the spin $2s$ contribution in $\Phi^{\rm biloc}|_{\eta_{(2s)}}$, we are picking out the largest spin harmonic from the spin $2s$ current block. This is a statement about the \emph{diagonal} entry of a triangular change of basis. The current-block label and the light-cone helicity label should not be identified before this triangular projection is performed: $\eta_{(2s')}$ denotes a conformal-current block, whereas $\Phi_{2s}^{(\pm)}$ denotes an angular, equivalently helicity, component of the bulk master field. The full answer for $\Phi_{2s}^{{\rm biloc}(+)}$ is schematically triangular,
\begin{equation}
\Phi_{2s}^{{\rm biloc}(+)}=\sum_{s'\ge s}\left(\Phi_{2s}^{{\rm biloc}(+)}\right)\big|_{\eta_{(2s')}} .
\end{equation}
The projection is \emph{upper triangular in spin} because higher-spin current blocks can source lower angular harmonics, but lower-spin blocks cannot source higher angular harmonics. In the formulas below we isolate the diagonal highest-harmonic contribution; the full helicity field is obtained only after the triangular sum over current blocks has been performed. To simplify the projection of $\Phi^{\rm biloc}\big|_{\eta^{(2s)}}$, it is convenient to introduce $\lambda \equiv 2s-r$. Using the standard relation between Jacobi and Gegenbauer polynomials,
\begin{equation}
P_r^{\left(\lambda-\frac12,\,\lambda-\frac12\right)}(x)=\frac{\rf{\lambda+\frac12}{r}}{\rf{2\lambda}{r}}
C_r^{\lambda}(x),\label{eq:jacobi-gegenbauer}
\end{equation}
and the Fourier expansion of $C_r^{\lambda}(\cos\theta)$,
\begin{equation}
C_r^{\lambda}(\cos\theta)=\sum_{k=0}^{r}\frac{\rf{\lambda}{k}\rf{\lambda}{r-k}}{k!(r-k)!}e^{i(r-2k)\theta},
\label{eq:gegenbauer-fourier}
\end{equation}
we obtain the explicit cosine series
\begin{equation}
P_r^{\left(\lambda-\frac12,\,\lambda-\frac12\right)}(\cos\theta)=\frac{\rf{\lambda+\frac12}{r}}{\rf{2\lambda}{r}}
\left[2\sum_{k=0}^{\lfloor (r-1)/2\rfloor}\frac{\rf{\lambda}{k}\rf{\lambda}{r-k}}{k!(r-k)!}\cos\bigl((r-2k)\theta\bigr)+
\delta_{r\,\mathrm{even}}\frac{\rf{\lambda}{r/2}^2}{\bigl((r/2)!\bigr)^2}\right].\label{eq:jacobi-cosine-series}
\end{equation}
Now combine \eqref{eq:jacobi-cosine-series} with the elementary Fourier expansion
\begin{equation}
(\sin\theta)^{\lambda}=\frac{1}{(2 i )^{\lambda}}\sum_{a=0}^{\lambda}(-1)^a\binom{\lambda}{a}
 e ^{ i (\lambda-2a)\theta}.\label{eq:sin-power-fourier}
\end{equation}
Multiplying \eqref{eq:jacobi-gegenbauer}, \eqref{eq:gegenbauer-fourier} and \eqref{eq:sin-power-fourier}, we find the finite exponential series
\begin{equation}
(\sin\theta)^{\lambda}P_r^{\left(\lambda-\frac12,\,\lambda-\frac12\right)}(\cos\theta)=
\frac{\rf{\lambda+\frac12}{r}}{\rf{2\lambda}{r}}\frac{1}{(2 i )^{\lambda}}\sum_{a=0}^{\lambda}\sum_{k=0}^{r}
(-1)^a\binom{\lambda}{a}\frac{\rf{\lambda}{k}\rf{\lambda}{r-k}}{k!(r-k)!}e^{2i(s-a-k)\theta}.
\label{eq:full-finite-fourier-series}
\end{equation}
Since $\lambda+r=2s$, every Fourier mode is of the form $ e ^{2 i  n\theta}$ with $n\in\{-s,-s+1,\dots,s\}$. Define
\begin{equation}
K_{2s,r}^{(\pm)}\equiv\frac{1}{2\pi}\int_0^{2\pi} d \theta\; e ^{\mp 2i s\theta}(\sin\theta)^{2s-r}
P_r^{\left(2s-r-\frac12,\,2s-r-\frac12\right)}(\cos\theta).\label{eq:I-def}
\end{equation}
Using \eqref{eq:full-finite-fourier-series}, the projection picks out a single Fourier mode and we find
\begin{align}
K_{2s,r}^{(+)}=\frac{1}{(2 i )^{\lambda}}\frac{\rf{\lambda+\frac12}{r}}{\rf{2\lambda}{r}}
\frac{\rf{\lambda}{r}}{r!},\qquad
K_{2s,r}^{(-)}=\frac{(-1)^{\lambda}}{(2 i )^{\lambda}}\frac{\rf{\lambda+\frac12}{r}}{\rf{2\lambda}{r}}
\frac{\rf{\lambda}{r}}{r!}.\label{eq:Iminus-first}
\end{align}
Now use the elementary identity\footnote{The identity
\begin{equation}
P_r^{(\lambda-\frac12,\lambda-\frac12)}(x)=\frac{(\lambda+\frac12)_r}{(2\lambda)_r}C_r^\lambda(x)
\label{eq:auto:0017}
\end{equation}
is singular at $\lambda=0$, but the singularity is removable. Using the generating function
\begin{equation}
\sum_{n\ge 0} C_n^\lambda(x)t^n=(1-2xt+t^2)^{-\lambda},
\label{eq:auto:0018}
\end{equation}
one finds, for $r\ge 1$ ($T_n(x)$ is the Chebyshev polynomial of the first kind $T_n(\cos\theta)=\cos(n\theta)$)
\begin{equation}
C_r^\lambda(x)=\frac{2\lambda}{r}T_r(x)+O(\lambda^2),
\label{eq:auto:0019}
\end{equation}
and therefore
\begin{equation}
\lim_{\lambda\to 0}\frac{(\lambda+\frac12)_r}{(2\lambda)_r}C_r^\lambda(x)=\frac{(\frac12)_r}{r!}T_r(x).
\qquad\Rightarrow\qquad P_r^{(-\frac12,-\frac12)}(x)=\frac{(\frac12)_r}{r!}T_r(x),
\label{eq:auto:0020}
\end{equation}
which is the correct endpoint formula used for $r=2s$.}
\begin{equation}
\frac{\rf{\lambda}{r}\rf{\lambda+\frac12}{r}}{\rf{2\lambda}{r}\,r!}=4^{-r}\binom{4s-1}{r},
\qquad \lambda=2s-r,\label{eq:key-pochhammer-identity}
\end{equation}
which gives the remarkably compact result
\begin{equation}
K_{2s,r}^{(\pm)}=2^{-2s-r}(\mp  i )^{2s-r}\binom{4s-1}{r}.\label{eq:I-final}
\end{equation}
Substituting \eqref{eq:I-final} into the projection of \eqref{eq:master-from-current-sector}, we obtain
\begin{align}
\frac{\left(\Phi_{2s}^{{\rm biloc}(\pm)}(X^+,P^+,X,Z)\right)\big|_{\eta_{(2s)}}}{Z}
%&=2^{3-2s}(2s)!\sum_{m=0}^{2s}\sum_{r=0}^{m}\binom{2s}{m}\binom{m}{r}
%(2 i )^r(- i )^{m-r}r!\frac{Z^{2s-r}}{(P^+)^{m}}\nonumber\\
%&\qquad\qquad\times K_{2s,r}^{(\pm)}\,\partial_X^{m-r}j_{2s}^{(m)}%(x^+,P^+;X)\nonumber\\
&=2^{3-4s}(2s)!\sum_{m=0}^{2s}\sum_{r=0}^{m}\binom{2s}{m}\binom{m}{r}\binom{4s-1}{r}2^{-r}(2i)^r
\nonumber\\
&\times (-i)^{m-r}(\mp i)^{2s-r}r!\frac{Z^{2s-r}}{(P^+)^{m}}\,\partial_X^{m-r}j_{2s}^{(m)}(x^+,P^+;X).\label{eq:phi-final-compact}
\end{align}
It is convenient to simplify the phase factors separately for the two helicities. For the `$+$' helicity we have $(2 i )^r(\!- i )^{2s-r}=2^r(-1)^{s+r}$, so that
\begin{align}
\Phi_{2s}^{{\rm biloc}(+)}(X^+,P^+,X,Z)\big|_{\eta_{(2s)}}&=2^{3-4s}(2s)!(-1)^s\sum_{m=0}^{2s}\sum_{r=0}^{m}
\binom{2s}{m}\binom{m}{r}\binom{4s-1}{r}(-1)^r(- i )^{m-r}\nonumber\\
&\qquad\qquad\qquad\times r!\frac{Z^{2s-r+1}}{(P^+)^{m}}\partial_X^{m-r}j_{2s}^{(m)}(x^+,P^+;X).\label{eq:phi-plus-final}
\end{align}
For the `$-$' helicity, $(2 i )^r(\!+ i )^{2s-r}=2^r(-1)^s$, so that
\begin{align}
\Phi_{2s}^{{\rm biloc}(-)}(X^+,P^+,X,Z)\big|_{\eta_{(2s)}}&=2^{3-4s}(2s)!(-1)^s\sum_{m=0}^{2s}\sum_{r=0}^{m}
\binom{2s}{m}\binom{m}{r}\binom{4s-1}{r}(-i)^{m-r}\nonumber\\
&\qquad\qquad\qquad\times r!\frac{Z^{2s-r+1}}{(P^+)^{m}}\partial_X^{m-r}j_{2s}^{(m)}(x^+,P^+;X).\label{eq:phi-minus-final}
\end{align}
Equations \eqref{eq:phi-plus-final} and \eqref{eq:phi-minus-final} give the diagonal, highest-harmonic contribution to the angularly projected helicity components, namely the contribution obtained from the spin-$2s$ current block itself. The full angular coefficient of the master field is triangular in the current spin label and receives additional contributions from all higher current blocks $\eta_{(2s')}$ with $s'\ge s$.

Our final formulas are more local than the standard HKLL expressions. Standard HKLL reconstructs a covariant bulk field from unconstrained local boundary data, and the reconstruction naturally appears as a smearing integral over a boundary region. Here we work with equal-$x^+$ bilocals and with completely gauge-fixed light-front helicity components, after current conservation has removed the $+$ polarizations. In this reduced description the nonlocality is encoded in the change of variables and in the descendant resummation, so the final answer takes the form of a finite differential operator evaluated at a kinematically determined boundary point. This suggests a deeper physical lesson: the standard smearing integral is the price one pays for keeping covariance manifest. Our formulas suggest that in the higher-spin/vector-model duality, once one gives up manifest covariance and works directly with the true propagating degrees of freedom, the dictionary may become dramatically more local in appearance. In Section~\ref{HKLLequivalence} we will check this interpretation by deriving the mixed HKLL kernel for the true propagating degrees of freedom and showing that it reproduces the same reduced master field.

\section{The equal $x^+$ bilocal obeys the bulk equation of motion} \label{sec:bulk-equation}

The bilocal reconstruction derived above has a striking feature: it yields a bulk field that is remarkably local in its dependence on the underlying bilocal CFT data. At first sight this may appear surprising. Precisely because of this tension, it is important to subject the reconstruction to independent consistency checks. In this section we provide such a check by demonstrating that the bulk field obtained from bilocal holography satisfies the correct bulk equation of motion and obeys the appropriate boundary conditions. These two properties are sufficiently constraining that, by standard uniqueness arguments, the reconstructed field must coincide with that obtained from other approaches to bulk reconstruction, most notably the HKLL prescription. In the following section we go beyond this structural argument and perform a detailed term-by-term comparison, explicitly verifying the agreement between the two constructions.

First we demonstrate that the equation of motion of the equal $x^+$ bilocal fluctuation implies the correct bulk light-front equation. The point is that the bilocal obeys a two-body light-front Schr\"odinger equation, and the change of variables diagonalizes the corresponding transverse kinetic operator exactly. The equal $x^+$ bilocal, in the mixed representation $\sigma(x^+,p_1^+,x_1,p_2^+,x_2)$ obeys the following equation of motion
\begin{equation}
i\partial_{x^+}\sigma=-\frac{1}{2p_1^+}\partial_{x_1}^2\sigma
-\frac{1}{2p_2^+}\partial_{x_2}^2\sigma.\label{eq:sec3-bilocal-two-body}
\end{equation}
This follows immediately from the free equations of motion of the fundamental fields. Writing $\sigma=\sigma_0+\eta$, the fluctuation obeys the same linear equation,
\begin{equation}
i\partial_{x^+}\eta=-\frac{1}{2p_1^+}\partial_{x_1}^2\eta
-\frac{1}{2p_2^+}\partial_{x_2}^2\eta.\label{eq:sec3-eta-two-body}
\end{equation}
Using the change of variables \eqref{coordtrans} and \eqref{invcoordtrans}, at fixed $P^+$ and fixed $\theta$, the chain rule gives
\begin{equation}
\frac{\partial X}{\partial x_1}=\cos^2\frac{\theta}{2},\qquad
\frac{\partial Z}{\partial x_1}=\sin\frac{\theta}{2}\cos\frac{\theta}{2},\qquad
\frac{\partial X}{\partial x_2}=\sin^2\frac{\theta}{2},\qquad
\frac{\partial Z}{\partial x_2}=-\sin\frac{\theta}{2}\cos\frac{\theta}{2}.
\label{eq:auto:0025}
\end{equation}
Therefore
\begin{equation}
\partial_{x_1}=\cos^2\frac{\theta}{2}\,\partial_X+
\sin\frac{\theta}{2}\cos\frac{\theta}{2}\,\partial_Z,\qquad
\partial_{x_2}=\sin^2\frac{\theta}{2}\,\partial_X-
\sin\frac{\theta}{2}\cos\frac{\theta}{2}\,\partial_Z.\label{eq:sec3-chain-rule}
\end{equation}
Using these two expressions and $P^+=p_1^++p_2^+$, we obtain the crucial identity
\begin{equation}
\frac{1}{p_1^+}\partial_{x_1}^2+\frac{1}{p_2^+}\partial_{x_2}^2=
\frac{1}{P^+}\left(\partial_X^2+\partial_Z^2\right).
\label{eq:sec3-kinetic-diagonalization}
\end{equation}
Substituting \eqref{eq:sec3-kinetic-diagonalization} into \eqref{eq:sec3-eta-two-body}, and using $X^+=x^+$, we find
\begin{equation}
i\partial_{X^+}\eta=-\frac{1}{2P^+}\left(\partial_X^2+\partial_Z^2\right)\eta.
\label{eq:sec3-eta-bulk-like}
\end{equation}
The master field is related to the bilocal fluctuation as dictated by \eqref{eq:Phi-eta-map}. The prefactor $4\pi\sqrt{p_1^+p_2^+}=2\pi P^+\sin\theta$ depends only on $P^+$ and $\theta$, so it commutes with $\partial_{X^+}$, $\partial_X$ and $\partial_Z$. Multiplying \eqref{eq:sec3-eta-bulk-like} by this prefactor gives
\begin{equation}
i\partial_{X^+}\Phi^{\rm biloc}=-\frac{1}{2P^+}\left(\partial_X^2+\partial_Z^2\right)\Phi^{\rm biloc}.
\label{eq:sec3-bulk-eom}
\end{equation}
This is precisely the free bulk light-front equation of motion. Since \eqref{eq:sec3-bulk-eom} contains no $\theta$-derivatives, it holds mode by mode in the angular decomposition of $\Phi$. Thus each angular, or helicity, component extracted from the equal $x^+$ bilocal master field obeys the above bulk equation of motion. This statement is slightly different from a statement about a single conformal-current block: as explained in Section~\ref{bilocalbulk}, the map from current blocks to helicity components is triangular. The equation of motion is therefore most naturally a statement about the reduced master field and its angular projections. The corresponding equation of motion for the individual spinning fields was obtained in~\cite{Metsaev:1999ui} in the bulk higher spin theory, after fixing the gauge and solving the constraint associated with this gauge choice.

Having established the correct bulk equation of motion, it remains to verify that the field reconstructed from bilocal holography has the correct asymptotic behaviour. This boundary condition is fixed by the GKPW~\cite{Gubser:1998bc,Witten:1998qj} prescription. A minor subtlety is that the standard GKPW dictionary is naturally formulated in de Donder gauge, whereas the bilocal construction produces bulk fields in light-cone gauge. The required gauge transformation from de Donder gauge to light-cone gauge was analyzed carefully by~\cite{Mintun:2014gua}, who showed that the light-cone fields obtained from the bilocal construction obey precisely the GKPW boundary condition. Thus the bilocal and HKLL reconstructions solve the same bulk equations with the same reduced boundary data, in the same light-front variables, once the same Lorentzian prescription is chosen. By uniqueness of the corresponding boundary-value problem, the two reconstructions must agree.

\section{HKLL reconstruction and an explicit equivalence proof}\label{HKLLequivalence}

In this section we prove that the bilocal reconstruction obtained in Sections~\ref{etaprimary} and \ref{bilocalbulk} agrees exactly with the reduced HKLL reconstruction. The logic has two ingredients. First, we derive the mixed light-front HKLL kernel explicitly and keep its radial phase throughout the comparison. Second, in the higher-spin sector we rewrite the HKLL coefficients in terms of the same current mode polynomial that underlies the bilocal derivation in Section~\ref{etaprimary}. The radial HKLL phase is then identified, on the two-particle support of the free vector model, with the relative transverse phase that separates the two bilocal endpoints. Only after this identification do we use the descendant resummation already established in Section~\ref{etaprimary}. Thus the argument is a concrete equivalence proof, rather than an appeal to uniqueness alone. Exactly as we did in our discussion of Section \ref{BBR}, we break the analysis up into sectors determined by the conserved current blocks, while remembering that the physical helicity fields are obtained only after the triangular angular projection described above.

\subsection{The mixed HKLL kernel}\label{subsec:mixed-hkll-kernel}

As discussed above, the reduced bulk field obeys the free light-front equation
\begin{equation}
\left(i\partial_{X^+}+\frac{1}{2P^+} (\partial_X^2+\partial_Z^2)\right)\Phi(X^+,P^+,X,Z,\theta)=0,\qquad Z>0.\label{eq:auto:0029}
\end{equation}
Fourier transforming in the boundary coordinates,
\begin{equation}
\widehat\Phi(\omega,P^+,k,Z;\theta)\equiv\int dX^+\,dX\;e^{i\omega X^++ikX}\Phi(X^+,P^+,X,Z,\theta),
\label{eq:sec3-double-fourier-field}
\end{equation}
turns \eqref{eq:sec3-bulk-eom} into
\begin{equation}
\left(\partial_Z^2+2P^+\omega-k^2\right)\widehat\Phi(\omega,P^+,k,Z;\theta)=0.\label{eq:sec3-ode-Z}
\end{equation}
Introduce
\begin{equation}
q(\omega,k)\equiv \sqrt{2P^+\omega-k^2+i0},\label{eq:sec3-q-def}
\end{equation}
with the branch fixed by analytic continuation from ${\rm Im}\,\omega>0$. The retarded solution is therefore
\begin{equation}
\widehat\Phi^{\rm HKLL}(\omega,P^+,k,Z;\theta)=e^{-iZq(\omega,k)} \,\widehat\Phi^{\rm HKLL}(\omega,P^+,k,0;\theta).\label{eq:sec3-fourier-solution}
\end{equation}
This is the mixed Fourier-space HKLL reconstruction formula. In position space we can rewrite this solution as a boundary convolution. Introduce the retarded Schr\"odinger kernel
\begin{equation}
G_{\rm ret}(\tau,P^+,\Delta X,Z)=\Theta(\tau)\,\frac{P^+}{2\pi i\,\tau}
\exp\left[\frac{iP^+}{2\tau}(\Delta X^2+Z^2)\right],\label{eq:sec3-Gret}
\end{equation}
where $\tau=X^+-x^+$ and $\Delta X=X-x$. The corresponding Poisson kernel is
\begin{equation}
K_e(\tau,P^+,\Delta X,Z)=\frac{1}{iP^+}\,\partial_ZG_{\rm ret}(\tau,P^+,\Delta X,Z)=\Theta(\tau)\,\frac{P^+Z}{2\pi i\,\tau^2}\exp\left[\frac{iP^+}{2\tau}(\Delta X^2+Z^2)\right].\label{eq:sec3-Ke-position}
\end{equation}
This can be derived explicitly by performing a Fourier transform. Define
\begin{equation}
\widetilde K_e(\omega,P^+,k,Z)\equiv \int_0^{\infty}d\tau\int_{-\infty}^{\infty}d(\Delta X)\;
e^{i\omega\tau-ik\Delta X}K_e(\tau,P^+,\Delta X,Z).\label{eq:sec3-Ke-tilde-def}
\end{equation}
Substituting \eqref{eq:sec3-Ke-position} into the above equation gives
\begin{align}
\widetilde K_e(\omega,P^+,k,Z)
&=\frac{P^+Z}{2\pi i}\int_0^\infty d\tau\,\tau^{-2}
e^{i\omega\tau+\frac{iP^+Z^2}{2\tau}}
\int_{-\infty}^{\infty}d(\Delta X)\,
\exp\!\left[-ik\Delta X+\frac{iP^+}{2\tau}\Delta X^2\right].\nonumber
\end{align}
For $\tau>0$ and $P^+>0$, the $\Delta X$ integral is the standard Fresnel Gaussian
\begin{equation}
\int_{-\infty}^{\infty}d(\Delta X)\,
\exp\!\left[-ik\Delta X+\frac{iP^+}{2\tau}\Delta X^2\right]=
e^{i\pi/4}\sqrt{\frac{2\pi\tau}{P^+}}\,
\exp\!\left[-\,\frac{i\tau k^2}{2P^+}\right].\label{eq:sec3-fresnel-step}
\end{equation}
so that, we obtain
\begin{equation}
\widetilde K_e(\omega,P^+,k,Z)=e^{-i\pi/4}Z\sqrt{\frac{P^+}{2\pi}}
\int_0^\infty d\tau\,\tau^{-3/2}
\exp\left[\frac{iP^+Z^2}{2\tau}+i\tau\!\left(\omega-\frac{k^2}{2P^+}\right)\right].
\label{eq:sec3-Ke-after-gaussian}
\end{equation}
To define the oscillatory $\tau$-integral unambiguously, insert a regulator $\epsilon>0$ and set
\begin{equation}
\alpha_\epsilon\equiv \epsilon-\frac{iP^+Z^2}{2},\qquad
\beta_\epsilon\equiv \epsilon-i\!\left(\omega-\frac{k^2}{2P^+}\right).\label{eq:sec3-alpha-beta}
\end{equation}
Then
\begin{equation}
\widetilde K_{e,\epsilon}(\omega,P^+,k,Z)
=
e^{-i\pi/4}Z\sqrt{\frac{P^+}{2\pi}}
\int_0^\infty d\tau\,\tau^{-3/2}e^{-\alpha_\epsilon/\tau-\beta_\epsilon\tau}.
\label{eq:sec3-Ke-regulated}
\end{equation}
For ${\rm Re}\,\alpha_\epsilon>0$ and ${\rm Re}\,\beta_\epsilon>0$, the standard Schwinger--Bessel integral gives
\begin{equation}
\int_0^\infty d\tau\,\tau^{-3/2}e^{-\alpha/\tau-\beta\tau}
=
\sqrt{\pi}\,\alpha^{-1/2}e^{-2\sqrt{\alpha\beta}},
\label{eq:sec3-schwinger-identity}
\end{equation}
so that
\begin{equation}
\widetilde K_{e,\epsilon}(\omega,P^+,k,Z)
=
e^{-i\pi/4}Z\sqrt{\frac{P^+}{2}}\,
\alpha_\epsilon^{-1/2}\,
e^{-2\sqrt{\alpha_\epsilon\beta_\epsilon}}.
\label{eq:sec3-Ke-regulated-evaluated}
\end{equation}
Now take $\epsilon\to 0^+$. Since $Z>0$ and $P^+>0$,
\begin{equation}
\alpha_\epsilon^{-1/2}\longrightarrow
e^{i\pi/4}\sqrt{\frac{2}{P^+}}\frac{1}{Z},\label{eq:sec3-alpha-limit}
\end{equation}
while
\begin{equation}
2\sqrt{\alpha_\epsilon\beta_\epsilon}\longrightarrow
iZ\sqrt{2P^+\omega-k^2+i0}=iZ\,q(\omega,k),\label{eq:sec3-beta-limit}
\end{equation}
with $q(\omega,k)$ defined in \eqref{eq:sec3-q-def}. Therefore
\begin{equation}
\widetilde K_e(\omega,P^+,k,Z)=e^{-iZq(\omega,k)}.\label{eq:sec3-kernel-transform}
\end{equation}
This is exactly the radial phase appearing in \eqref{eq:sec3-fourier-solution}. Hence the Fourier-space solution and the position-space smearing formula are equivalent, and we may write
\begin{equation}
\Phi^{\rm HKLL}\big|_{\eta^{(2s)}}(X^+,P^+,X,Z,\theta)=\int_0^\infty d\tau\int_{-\infty}^{\infty}dx\;
K_e(\tau,P^+,X-x,Z)\,\mathcal B_{2s}(X^+-\tau,P^+,x;\theta),\label{eq:sec3-hkll-smearing}
\end{equation}
where the boundary condition is
\begin{equation}
\mathcal B_{2s}(X^+,P^+,X;\theta)=\Phi^{\rm HKLL}\big|_{\eta^{(2s)}}(X^+,P^+,X,0,\theta).
\label{eq:sec3-boundary-data-def}
\end{equation}
Here we have broken the reconstructed bulk field into sectors determined by the current blocks. The boundary condition $\mathcal B_{2s}(X^+,P^+,X;\theta)$ will only receive contributions from the spin-$2s$ primary. Appendix~\ref{poissonkernel} reviews the general Poisson-kernel philosophy in a simpler setting.

\subsection{The correct boundary condition}\label{subsec:correct-boundary-data}

We now determine \eqref{eq:sec3-boundary-data-def} directly from the bilocal answer. For $s\ge 1$, take the $Z\to 0$ limit of equation \eqref{eq:master-from-current-sector}. Since every term with $r<2s$ carries a positive power of $Z$, only the term $r=2s$ survives. Since $r\le m\le 2s$, this also forces $m=2s$. Therefore
\begin{equation}
\Phi^{\rm biloc}\big|_{\eta^{(2s)}}(X^+,P^+,X,0,\theta)=16\pi\,(2s)!\frac{(i)^{2s}(2s)!}{(P^+)^{2s}}
P_{2s}^{\left(-\frac12,-\frac12\right)}(\cos\theta)\,j_{2s}^{(2s)}(X^+,P^+,X).
\label{eq:sec3-boundary-limit-raw}
\end{equation}
Using
\begin{equation}
P_{2s}^{\left(-\frac12,-\frac12\right)}(\cos\theta)=\frac{\left(\frac12\right)_{2s}}{(2s)!}\,T_{2s}(\cos\theta)
=\frac{\Gamma\left(2s+\frac12\right)}{\sqrt\pi\,(2s)!}\cos(2s\theta)
\label{eq:sec3-endpoint-jacobi}
\end{equation}
and $i^{2s}=(-1)^s$, we obtain
\begin{equation}
\mathcal B_{2s}(X^+,P^+,X;\theta)=16\pi(-1)^s\frac{(2s)!\,\Gamma\left(2s+\frac12\right)}{\sqrt\pi\,(P^+)^{2s}}
\cos(2s\theta)\,j_{2s}^{(2s)}(X^+,P^+,X),
\qquad s\ge 1.
\label{eq:sec3-physical-B-profile}
\end{equation}
Thus the boundary condition is set by the light-cone current $j_{2s}^{(2s)}$, multiplied by the angular factor $\cos(2s\theta)$ and the normalization is dictated by the bilocal boundary limit. Up to an overall normalization, this result agrees with that of~\cite{Mintun:2014gua}. We have nevertheless derived it directly from the bilocal bulk field in order to ensure that the normalization of boundary conditions is treated consistently in the two reconstruction frameworks we aim to compare.

The scalar sector is treated separately. We distinguish the local scalar primary $j_0^{(0)}=:\phi^a\phi^a:$ from the projected and resummed scalar conformal-block, denoted $\mathcal J_0^{\rm blk}$. From the scalar contribution in \eqref{eq:eta-xspace-final} and the map \eqref{eq:Phi-eta-map}, we get
\begin{eqnarray}
\Phi^{\rm biloc}\big|_{\eta^{(0)}}(X^+,P^+,X,Z,\theta)&=&2\pi P^+\sin\theta\;\frac{4}{P^+\sqrt{1-\nu^2}}\,\mathcal J_0^{\rm blk}(X^+,P^+,X)\cr\cr
&=&8\pi\,\mathcal J_0^{\rm blk}(X^+,P^+,X),
\label{eq:sec3-scalar-boundary-value}
\end{eqnarray}
where we recall that $\nu=\cos\theta$ so that $\sqrt{1-\nu^2}=\sin\theta$. Hence
\begin{equation}
\mathcal B_0(X^+,P^+,X;\theta)=8\pi\,\mathcal J_0^{\rm blk}(X^+,P^+,X).
\label{eq:sec3-scalar-boundary-data}
\end{equation}

\subsection{Direct comparison in the scalar sector}\label{subsec:scalar-comparison}

We first verify the scalar sector directly. Again we distinguish the projected and resummed scalar conformal block $\mathcal J_0^{\rm blk}$, and the local scalar primary $j_0^{(0)}$ itself. Fourier transforming the boundary field in \eqref{eq:sec3-scalar-boundary-data} gives
\begin{equation}
\Phi^{\rm HKLL}\big|_{\eta^{(0)}}(X^+,P^+,X,Z,\theta)
=
8\pi\int \frac{d\omega\,dk}{(2\pi)^2}\,e^{-i\omega X^+-ikX}\,
e^{-iZq(\omega,k)}\,
\widehat {\mathcal J}_{0}^{\rm blk}(\omega,P^+,k).
\label{eq:sec3-scalar-hkll-fourier}
\end{equation}
On the other hand, the bilocal formula \eqref{eq:sec3-scalar-boundary-value} is independent of $Z$. The point which must be kept explicit is that this statement is not a statement about the bare local primary $j_0^{(0)}$.  It is a statement about the scalar conformal block of the bilocal after the descendant tower has been resummed. The block decomposition is a decomposition into independent conformal modules, while the light-front evolution of the bilocal is linear and commutes with the conformal action. Hence the scalar block may be inserted separately into the bulk equation \eqref{eq:sec3-bulk-eom}. Since the scalar-block contribution has no $Z$-dependence, this gives
\begin{equation}
\left(i\partial_{X^+}+\frac{1}{2P^+}\partial_X^2\right)\mathcal J_0^{\rm blk}(X^+,P^+,X)=0.
\label{eq:sec3-scalar-boundary-eom}
\end{equation}
Equivalently, this equation is the center-of-mass Schr\"odinger equation obeyed by the projected and resummed two-particle scalar block. It should not be read as a free equation for $j_0^{(0)}$ itself. Fourier transforming \eqref{eq:sec3-scalar-boundary-eom} gives
\begin{equation}
\left(2P^+\omega-k^2\right)\widehat {\mathcal J}_0^{\rm blk}(\omega,P^+,k)=0.
\label{eq:sec3-scalar-onshell}
\end{equation}
Thus $\widehat {\mathcal J}_0^{\rm blk}$ is supported on\footnote{See Appendix \ref{scalarsupport} for further discussion.} $2P^+\omega-k^2=0$, and on that support
\begin{equation}
q(\omega,k)=\sqrt{2P^+\omega-k^2+i0}=0.
\label{eq:sec3-scalar-qzero}
\end{equation}
Therefore the radial phase in \eqref{eq:sec3-scalar-hkll-fourier} collapses to $1$ and we obtain
\begin{equation}
\Phi^{\rm HKLL}\big|_{\eta^{(0)}}(X^+,P^+,X,Z,\theta)=8\pi\,\mathcal J_0^{\rm blk}(X^+,P^+,X).
\label{eq:sec3-scalar-hkll-final}
\end{equation}
Comparing \eqref{eq:sec3-scalar-hkll-final} with \eqref{eq:sec3-scalar-boundary-value}, we conclude that the scalar HKLL and bilocal reconstructions agree exactly.

\subsection{Higher-spin comparison for $s\ge 1$}\label{subsec:higher-spin-comparison}

We now turn to the higher-spin sector. We rewrite the HKLL coefficients in terms of the same current mode polynomial that appeared in the bilocal derivation. Once that common polynomial is identified, the descendant resummation of Section~\ref{etaprimary} applies verbatim, and the remaining step is the explicit inverse Fourier transform in $(\omega,k)$. To isolate the contributions associated with the monomials $(y^-)^m y^{2s-m}$, we use $y^-$ and $y$ only as bookkeeping variables. We define
\begin{equation}
\Xi_{s,m}^{\rm HKLL}(X^+,P^+,X,Z;\theta)\equiv\binom{2s}{m}^{-1}
\bigl[(y^-)^m y^{2s-m}\bigr]\,\Phi^{\rm HKLL}\big|_{\eta^{(2s)}},
\label{eq:sec3-Xi-def-coeff}
\end{equation}
where $\bigl[(y^-)^m y^{2s-m}\bigr]$ denotes coefficient extraction in the same formal expansion used in Section~\ref{etaprimary} to organize the reduced current components.
The key point is that these coefficients can be computed exactly from the HKLL reconstruction without dropping the radial phase. Appendix~\ref{identity} keeps the factor $e^{-iZq(\omega,k)}$ until the two-particle support of the bilocal is imposed. On that support $q(\omega,k)$ becomes the relative transverse momentum, and the HKLL factor $e^{-ikX}e^{-iZq(\omega,k)}$ becomes precisely the phase $e^{-i(p_1x_1+p_2x_2)}$ of the separated bilocal endpoints. Thus radial propagation in HKLL is converted into endpoint separation in the bilocal representation. After this identification the HKLL integrand is rewritten in terms of the common mode polynomial and then reduced, using the Section~\ref{etaprimary} resummation, to an explicit coefficient formula. The result is
\begin{align}
\Xi_{s,m}^{\rm HKLL}
&=
2\pi\,2^{3-2s}(2s)!
\sum_{r=0}^{m}\binom{m}{r}(2i)^r(-i)^{m-r}r!\,
\frac{Z^{2s-r}}{(P^+)^{m}}
(\sin\theta)^{2s-r}
\nonumber\\[1mm]
&\qquad\qquad\times
P_r^{\left(2s-r-\frac12,\;2s-r-\frac12\right)}(\cos\theta)\,
\partial_X^{m-r}j_{2s}^{(m)}(X^+,P^+,X).
\label{eq:sec3-hkll-coeff-explicit}
\end{align}
Summing over $m$ with the same bookkeeping basis used in Section~\ref{etaprimary}, we recover
\begin{align}
\Phi^{\rm HKLL}\big|_{\eta^{(2s)}}(X^+,P^+,X,Z,\theta)
&=
\frac{16\pi(2s)!}{2^{2s}}
\sum_{m=0}^{2s}\sum_{r=0}^{m}\binom{2s}{m}\binom{m}{r}
(2i)^r(-i)^{m-r}r!\,
\frac{Z^{2s-r}}{(P^+)^{m}}
\nonumber\\[1mm]
&\!\!\!\!\!\!\!\!\times
(\sin\theta)^{2s-r}
P_r^{\left(2s-r-\frac12,\;2s-r-\frac12\right)}(\cos\theta)\,
\partial_X^{m-r}j_{2s}^{(m)}(X^+,P^+,X).
\label{eq:sec3-hkll-master-expanded}
\end{align}
This is identical, term by term, to the bilocal formula \eqref{eq:master-from-current-sector}. Therefore
\begin{equation}
\Phi^{\rm HKLL}\big|_{\eta^{(2s)}}(X^+,P^+,X,Z,\theta)=
\Phi^{\rm biloc}\big|_{\eta^{(2s)}}(X^+,P^+,X,Z,\theta),
\qquad s\ge 1.
\label{eq:sec3-master-field-equality}
\end{equation}
Together with the scalar comparison in \eqref{eq:sec3-scalar-hkll-final}, this proves the equality of the full reduced master fields for all current blocks. Once \eqref{eq:sec3-master-field-equality} is established at fixed $\theta$, angular projection may be applied to both sides. The diagonal highest-harmonic projection of the spin-$2s$ current block reproduces \eqref{eq:phi-plus-final} and \eqref{eq:phi-minus-final}; the full physical helicity field is then obtained by the triangular sum over all current blocks $\eta_{(2s')}$ with $s'\geq s$.

\section{Subregion reconstruction from bilocal holography and modular organization near the edge}\label{sec:subregion-modular}

In Section~\ref{HKLLequivalence} we proved the exact equality of the reduced HKLL and bilocal reconstructions at fixed $\theta$ and fixed current block,
\begin{equation}
\Phi^{\mathrm{HKLL}}\big|_{\eta^{(2s)}}(X^+,P^+,X,Z,\theta)=
\Phi^{\mathrm{biloc}}\big|_{\eta^{(2s)}}(X^+,P^+,X,Z,\theta)\qquad \text{for all } s\geq 0.
\label{eq:sec6-global-equality}
\end{equation}
The natural next question is what this implies for subregion reconstruction and for the organization of operators near an entangling surface. The discussion of this section has three logically distinct themes. First, there is an \emph{exact kinematical layer}.  On the equal-$X^+$ slice, the bilocal endpoint map \eqref{eq:sec6-endpoint-map} gives an exact criterion for when a bulk point lies in the AdS--Rindler wedge associated with a boundary interval.  This implies an exact subregion statement for the reduced master field with the auxiliary angular variable $\theta$ retained. Second, there is an \emph{exact reduced flow on endpoint space}.  The interval condition singles out a canonical one-dimensional conformal flow on the endpoint line, namely the unique M\"obius flow preserving the endpoints $\pm \ell$.  This flow organizes the edge limit\footnote{Modular flow becomes simplest and most universal near the edge for an interval. Exact modular flow is a conformal transformation preserving $\pm\ell$; close to either endpoint it reduces to a local Rindler boost. This is the reason why the near edge region is conceptually powerful -- it isolates the universal short distance structure of modular flow independent of the detailed global geometry of the interval. Further, bilocals are naturally adapted to this limit. A bilocal $x_1,x_2$ transforms under modular flow by transporting each endpoint. Near edge-endpoint action is multiplicative so that bilocal operators are expected to be objects with simple modular frequency.} exactly and leads to the exact coordinates $(\varrho,\upsilon)$ in which the endpoint dynamics becomes translation in $\varrho$. Third, there is the passage to the \emph{mixed} $(p_i^+,x_i)$ representation on the null plane.  Here the modular question changes character.  Because we work on $X^+=0$, the underlying bilocal lives on a null hypersurface, and the mixed representation is obtained by Fourier transform in $x_i^-$.  As a result, a local rescaling of the null coordinates becomes a dilation in $p_i^+$.  This means that exact translation in $\varrho$ for the reduced endpoint variables does \emph{not} by itself imply pure translation at fixed $p_i^+$.  Instead, one is naturally led to a Mellin basis in $p_i^+$, and only after isolating the universal momentum-dilation piece does one recover a clean edge translation problem.  The approximate modular-frequency statements below are formulated precisely in that language.

\subsection{The exact wedge statement from the endpoint map}\label{subsec:exact-wedge-statement}

Fix the boundary interval
\begin{equation}
A=(-\ell,\ell),\qquad \ell>0.\label{eq:sec6-interval-A}
\end{equation}
Its associated AdS--Rindler wedge on a fixed-$X^+$, fixed-$P^+$ slice is
\begin{equation}
W_A=\bigl\{(X,Z): Z>0,\; X^2+Z^2<\ell^2\bigr\}.\label{eq:sec6-wedge-define}
\end{equation}
The bilocal-to-bulk map on this slice is
\begin{equation}
x_1=X+Z\tan\frac\theta2,\qquad x_2=X-Z\cot\frac\theta2,\qquad Z>0,\qquad 0<\theta<\pi.
\label{eq:sec6-endpoint-map}
\end{equation}
To reconstruct from bilocals supported in $A$, both endpoints must lie in $A$ which implies
\begin{equation}
-\ell<X+Z\tan\frac\theta2<\ell,\qquad -\ell<X-Z\cot\frac\theta2<\ell.
\label{eq:sec6-endpoint-inequalities}
\end{equation}
Since $0<\theta<\pi$, both $\tan(\theta/2)$ and $\cot(\theta/2)$ are positive. Solving \eqref{eq:sec6-endpoint-inequalities} therefore gives
\begin{equation}
2\arctan\Bigl(\frac{Z}{\ell+X}\Bigr)<\theta<2\arctan\Bigl(\frac{\ell-X}{Z}\Bigr).
\label{eq:sec6-theta-range}
\end{equation}
This interval is non-empty precisely when
\begin{equation}
\frac{Z}{\ell+X}<\frac{\ell-X}{Z}\qquad\Longleftrightarrow\qquad X^2+Z^2<\ell^2.
\label{eq:sec6-wedge-criterion}
\end{equation}
Thus the endpoint map gives the exact criterion
\begin{equation*}
\exists\,\theta\in(0,\pi)\ \text{with}\ x_1,x_2\in A\qquad\Longleftrightarrow\qquad (X,Z)\in W_A.
\end{equation*}

Let $\mathcal B_A$ denote the algebra generated by equal-$X^+$ bilocals whose endpoints lie in $A$.  Then for any $(X,Z)\in W_A$ and any $\theta$ satisfying \eqref{eq:sec6-theta-range}, the reduced master field
\begin{equation}
\Phi^{\rm biloc}\big|_{\eta^{(2s)}}(X^+,P^+,X,Z,\theta)=2\pi P^+\sin\theta\,\eta^{(2s)}(x^+,p_1^+,x_1,p_2^+,x_2) \label{eq:sec6-fixed-theta-master-in-algebra}
\end{equation}
belongs to $\mathcal B_A$. Combining this with the exact bilocal/HKLL equality proved in Section~\ref{HKLLequivalence},
\begin{equation}
\Phi^{\mathrm{biloc}}\big|_{\eta^{(2s)}}(X^+,P^+,X,Z,\theta)=\Phi^{\mathrm{HKLL}}\big|_{\eta^{(2s)}}(X^+,P^+,X,Z,\theta),\label{eq:sec6-fixed-theta-equality}
\end{equation}
we conclude that for every $(X,Z)\in W_A$ and every allowed $\theta$ in the range \eqref{eq:sec6-theta-range}, the reduced bulk master field $\Phi^{\rm biloc}\big|_{\eta^{(2s)}}(X^+,P^+,X,Z,\theta)$ has an exact representative in the interval bilocal algebra $\mathcal B_A$.

This is the exact subregion statement implied by bilocal holography.  It is exact, but it is at the level of the \emph{reduced master field with $\theta$ retained}.  It does \emph{not} immediately imply a corresponding exact statement for the angularly projected helicity fields $\Phi^{{\rm biloc}(\pm)}_{2s}(X^+,P^+,X,Z)$, because once one restricts to the subregion only a limited $\theta$-range is available at fixed $(X,Z)$. The full angular projection \eqref{eq:projection-formula} is no longer a purely local operation inside that restricted range.  This indicates that a distinction must be made between exact kinematics for the master field and more delicate statements for projected fixed-spin sectors.

\subsection{Exact modular coordinates and the edge limit}\label{subsec:exact-modular-coordinates}

In this section we seek further evidence for the bilocal reconstruction formula from modular flow. The relevant point of comparison is the JLMS relation~\cite{Jafferis:2015del}, which identifies the boundary modular Hamiltonian of a subregion with the corresponding bulk modular Hamiltonian in the entanglement wedge, together with the appropriate area operator. Equivalently, within the semiclassical code subspace, boundary modular flow is represented in the bulk by the modular flow of the entanglement wedge. This observation has played a central role in the modern understanding of entanglement wedge reconstruction and suggests a natural test of the bilocal reconstruction: the bulk region accessed by a boundary subregion should be the region naturally generated by the corresponding modular evolution. Exact modular flows, however, are difficult to construct in general; apart from highly symmetric cases, explicit closed-form expressions are rare. We shall therefore adopt a more modest strategy. We identify natural candidate flows with the qualitative features expected of modular evolution and ask what they teach us about the localization properties of the bilocal reconstruction. In this sense the analysis below should be viewed not as a derivation from an exact modular Hamiltonian, but as a modular-flow-inspired consistency check of the reconstruction formula.

We will now consider flows on the endpoint space. We isolate the canonical one-dimensional structure singled out by the interval, without making a claim about the full modular operator of the null-plane algebra. The subregion condition is imposed directly on the endpoint coordinates $x_1$ and $x_2$. There is a distinguished organization of the geometry of the endpoints in terms of the unique one-parameter conformal flow on the line that preserves the interval and fixes its two boundary points.  This is the M\"obius subgroup of $PSL(2,\mathbb R)$ preserving $\pm \ell$. Further, there is also a bulk-geometric reason for this choice which is intrinsic to bilocal holography. On the equal-$X^+$ bulk slice the metric is
\begin{equation}
ds^2=\frac{dX^2+dZ^2}{Z^2},\label{eq:sec6-H2-metric}
\end{equation}
so the slice is the hyperbolic plane $H_2$.  The wedge $W_A$ is bounded by the geodesic $X^2+Z^2=\ell^2$ anchored at the two boundary points $\pm\ell$.  The hyperbolic plane has a unique one-parameter isometry subgroup that preserves this geodesic and its two boundary endpoints.  On the boundary line, this isometry acts precisely by the M\"obius flow preserving $(-\ell,\ell)$.  Thus the one-dimensional flow we consider is not arbitrary: it is the boundary action of the canonical $H_2$ isometry\footnote{The $H_2$ isometry that preserves the wedge $W_A$ is generated by the vector field $$\xi_A=\frac{\pi}{\ell}\left[(\ell^2-X^2-Z^2)\partial_X-2XZ\partial_Z\right].$$} singled out by the wedge geometry of the equal-$X^+$ slice itself.

Consider endpoint distances from the right entangling point,
\begin{equation}
y_1\equiv \ell-x_1,\qquad y_2\equiv \ell-x_2.\label{eq:auto:0030}
\end{equation}
The right-edge limit is obtained by taking $y_1/\ell\ll1$ and $y_2/\ell\ll1$. Let $f_\tau(x)$ denote the unique M\"obius transformation preserving $(-\ell,\ell)$ and fixing $\pm\ell$.  It is convenient to characterize it by the cross-ratio type variable
\begin{equation}
u=\frac{\ell-x}{\ell+x},
\end{equation}
which maps $(-\ell,\ell)$ to the positive half-line.  The canonical flow acts on the half-line by dilation, $u\mapsto u_\tau=e^{-2\pi\tau}u$. Pulling this back to the endpoint line yields the finite transformation
\begin{equation}
\frac{\ell-f_\tau(x)}{\ell+f_\tau(x)}=e^{-2\pi\tau}\frac{\ell-x}{\ell+x}.
\end{equation}
Solving for $f_\tau(x)$ gives the familiar M\"obius form
\begin{equation}
x(\tau)=\ell\,\frac{\frac{x}{\ell}+\tanh(\pi\tau)}{1+\frac{x}{\ell}\tanh(\pi\tau)}.\label{eq:auto:0032}
\end{equation}
Differentiating at $\tau=0$ yields the corresponding vector field,
\begin{equation}
\frac{dx(\tau)}{d\tau}=\frac{\pi}{\ell}\bigl(\ell^2-x(\tau)^2\bigr).
\label{eq:auto:0031}
\end{equation}
This equation is best understood geometrically.  The factor $\ell^2-x^2$ forces the generator to vanish at the endpoints $x=\pm\ell$, so the entangling points are fixed.  Inside the interval, $\ell^2-x^2>0$, so positive flow time moves points monotonically to the right while keeping them in the interval.  It is therefore the canonical endpoint-preserving flow on the reduced endpoint space. Equivalently, in terms of $y=\ell-x$, one has
\begin{equation}
y_\tau=\frac{e^{-2\pi\tau}y}{1-\frac{1-e^{-2\pi\tau}}{2\ell}\,y}.
\label{eq:exact-y-flow}
\end{equation}
This form makes the edge behavior clear.  Near the right edge $y/\ell\ll1$, and \eqref{eq:exact-y-flow} becomes
\begin{equation}
y_\tau=e^{-2\pi\tau}y+O\!\left(\frac{y^2}{\ell}\right).
\end{equation}
To leading order, the distance to the right entangling point is rescaled by $e^{-2\pi\tau}$.  This is the edge analogue of the familiar boost behavior near a horizon. The exact linearizing variable is
\begin{equation}
u\equiv \frac{y}{2\ell-y}=\frac{\ell-x}{\ell+x}.\label{eq:u-def}
\end{equation}
This variable is the coordinate adapted to the interval endpoints.  It vanishes at the right endpoint, diverges at the left endpoint, and converts the nonlinear fractional action \eqref{eq:exact-y-flow} into simple multiplication.  Indeed, using \eqref{eq:exact-y-flow},
\begin{equation}
2\ell-y_\tau=\frac{2\ell-y}{1-\frac{1-e^{-2\pi\tau}}{2\ell}\,y},\qquad\Rightarrow\qquad
u_\tau=\frac{y_\tau}{2\ell-y_\tau}=e^{-2\pi\tau}\frac{y}{2\ell-y}=e^{-2\pi\tau}u.\label{eq:u-linear-flow}
\end{equation}
Applying this separately to the two endpoints gives
\begin{equation}
u_{1,\tau}=e^{-2\pi\tau}u_1,\qquad u_{2,\tau}=e^{-2\pi\tau}u_2,\qquad u_i=\frac{\ell-x_i}{\ell+x_i}.\label{eq:u12-flow}
\end{equation}
This suggests the exact edge-adapted coordinates
\begin{equation}
\varrho=-\frac{1}{4\pi}\log(u_1u_2),\qquad\upsilon=\frac{1}{4\pi}\log\frac{u_2}{u_1}.\label{eq:exact-mod-coords}
\end{equation}
The variable $\varrho$ measures the overall proximity of the pair of endpoints to the right edge: when both $u_1$ and $u_2$ are small, $\varrho$ is large.  The variable $\upsilon$ measures the relative position of the two endpoints once the common scale has been factored out.  In this precise sense, $\varrho$ is the ``radial'' variable of the endpoint pair under the canonical flow, while $\upsilon$ labels the orbit. In terms of the endpoints themselves,
\begin{equation}
\varrho=-\frac{1}{4\pi}\log\left[\frac{(\ell-x_1)(\ell-x_2)}{(\ell+x_1)(\ell+x_2)}\right],\qquad
\upsilon=\frac{1}{4\pi}\log\left[\frac{(\ell-x_2)(\ell+x_1)}{(\ell-x_1)(\ell+x_2)}\right].
\label{eq:exact-mod-coords-endpoints}
\end{equation}
Using \eqref{eq:u12-flow}, we have
\begin{equation}
\varrho_\tau=-\frac{1}{4\pi}\log(u_{1,\tau}u_{2,\tau})=-\frac{1}{4\pi}\log\!\bigl(e^{-4\pi\tau}u_1u_2\bigr)=
\varrho+\tau,\label{eq:varrho-translation}
\end{equation}
while
\begin{equation}
\upsilon_\tau=\frac{1}{4\pi}\log\frac{u_{2,\tau}}{u_{1,\tau}}=\frac{1}{4\pi}\log\frac{u_2}{u_1}=\upsilon.
\label{eq:upsilon-invariant}
\end{equation}
Thus the canonical endpoint-preserving flow acts as translation in $\varrho$, with $\upsilon$ an orbit label. It is useful to express the exact orbit label $\upsilon$ in bulk-adapted variables. Introduce the boundary midpoint and separation,
\begin{equation}
x=\frac{x_1+x_2}{2},\qquad y=\frac{x_1-x_2}{2},\label{eq:auto:0036}
\end{equation}
so that
\begin{equation}
x=X-Z\cot\theta,\qquad y=\frac{Z}{\sin\theta}. \label{eq:auto:0037}
\end{equation}
From \eqref{eq:exact-mod-coords-endpoints} one finds
\begin{equation}
\tanh(2\pi\upsilon)=\frac{u_2-u_1}{u_2+u_1}=\frac{2\ell\,y}{\ell^2-x^2+y^2}.\label{eq:upsilon-xy-exact}
\end{equation}
Using the bilocal-to-bulk map then gives the exact relation
\begin{equation}
\tanh(2\pi\upsilon)=\frac{2\ell Z}{\bigl(\ell^2-X^2+Z^2\bigr)\sin\theta+2XZ\cos\theta}.
\label{eq:upsilon-theta-exact}
\end{equation}
Thus, for fixed bulk point $(X,Z)$, the auxiliary angle $\theta$ may be traded for the exact orbit label $\upsilon$ on any branch for which $\partial_\theta\upsilon\neq0$.  This relation sheds light on how the fixed-spin projection behaves near the edge.

Equations \eqref{eq:auto:0030}--\eqref{eq:upsilon-theta-exact} are exact.  They are exact because they are statements about the reduced endpoint geometry of the interval and the exact bilocal-to-bulk map.  What they do \emph{not} yet tell us is how the full null-plane bilocal in the mixed $(p_i^+,x_i)$ representation transforms under the true modular flow of the interval algebra. We take up this issue in the next subsection.

\subsection{Passage to the mixed representation: a natural modular lift}\label{subsec:mixed-representation-assumption}

The equal-$x^+$ bilocal lives on a null plane. Its mixed representation is obtained by Fourier transforming in the null coordinates $x_i^-$.  To consider the modular flow of the actual bilocal operator, we must understand not only how the endpoint labels $x_i$ move, but also how the null coordinates $x_i^-$ transform. The reduced flow derived in Section~\ref{subsec:exact-modular-coordinates} acts only on the endpoint line. We need a lift of that flow to the null plane.  There is a natural minimal choice.  Let $f_\tau(x)$ be the finite M\"obius map \eqref{eq:auto:0032}.  The corresponding local rescaling of the null generators is
\begin{equation}
x_i\longmapsto f_\tau(x_i),\qquad x_i^-\longmapsto f_\tau'(x_i)\,x_i^-.
\label{eq:sec6-candidate-null-lift}
\end{equation}
This is the simplest lift compatible with two basic requirements: the transverse coordinate should transform by the canonical endpoint-preserving M\"obius map, and the ambient null-plane conformal structure should be respected.  It is also suggested by known modular-flow results on null planes, where the flow acts locally along the null generators rather than purely in the transverse direction.\footnote{For the null-plane modular-flow picture in higher-dimensional CFT, see~\cite{Casini:2017roe}.  The present use of \eqref{eq:sec6-candidate-null-lift} should be understood as the natural local lift of the exact reduced endpoint flow, not as an independent theorem about the full modular operator of our bilocal interval algebra.} Infinitesimally, writing
\begin{equation}
f_\tau(x)=x+\tau\,\zeta(x)+O(\tau^2),\qquad\zeta(x)=\frac{\pi}{\ell}\bigl(\ell^2-x^2\bigr),
\end{equation}
we find
\begin{equation}
\delta x_i=\zeta(x_i),\qquad \delta x_i^-=\zeta'(x_i)\,x_i^-, \qquad \zeta'(x)=-\frac{2\pi x}{\ell}.
\label{eq:sec6-candidate-infinitesimal}
\end{equation}
Thus the corresponding candidate differential operator on endpoint position space is
\begin{equation}
\mathcal D^{\mathrm{cand}}_{A,\mathrm{pos}}=\sum_{i=1}^{2}\left[\zeta(x_i)\partial_{x_i}+\zeta'(x_i)x_i^-\partial_{x_i^-}+\frac12\zeta'(x_i)\right].
\label{eq:sec6-candidate-pos-generator}
\end{equation}
We now pass to the mixed representation.  Write
\begin{equation}
\eta(x^+,p_1^+,x_1,p_2^+,x_2)=\int dx_1^-\,dx_2^-\,e^{i p_1^+x_1^-+i p_2^+x_2^-}\,\widetilde\eta(x^+,x_1^-,x_1,x_2^-,x_2).\label{eq:sec6-mixed-fourier-definition}
\end{equation}
The crucial identity is that a dilation in $x_i^-$ becomes a dilation in $p_i^+$, since
\begin{equation}
\int dx^-\,e^{ip^+x^-}\,x^-\partial_{x^-}\widetilde\eta=-\bigl(1+p^+\partial_{p^+}\bigr)\eta.
\label{eq:sec6-fourier-dilation-identity}
\end{equation}
Therefore \eqref{eq:sec6-candidate-pos-generator} becomes the following candidate differential operator in the mixed representation:
\begin{equation}
\mathcal D^{\mathrm{cand}}_{A,\mathrm{mix}}=\sum_{i=1}^{2}\left[\zeta(x_i)\partial_{x_i}-\zeta'(x_i)\bigl(1+p_i^+\partial_{p_i^+}\bigr)+\frac12\zeta'(x_i)\right].
\label{eq:sec6-candidate-mixed-generator}
\end{equation}
In the mixed $(p_i^+,x_i)$ representation, the modular problem is no longer pure transport in the endpoint coordinates.  Instead, the same local conformal rescaling that moves the endpoints also dilates the lightfront momenta. It is useful to rewrite the candidate generator in the exact endpoint variables $(\varrho,\upsilon)$.  Since the reduced flow acts as $\varrho\mapsto \varrho+\tau$, $\upsilon\mapsto\upsilon$, the endpoint part of the generator is $\partial_\varrho$.  Thus
\begin{equation}
\mathcal D^{\mathrm{cand}}_{A,\mathrm{mix}}=\partial_\varrho-\sum_{i=1}^{2}\zeta'\bigl(x_i(\varrho,\upsilon)\bigr)\bigl(1+p_i^+\partial_{p_i^+}\bigr)+\sum_{i=1}^2\frac12\zeta'(x_i)\label{eq:sec6-candidate-mixed-generator-varrho}
\end{equation}
The edge limit is now completely transparent:  since $x_i=\ell-y_i$ and $y_i=O(e^{-2\pi\varrho})$ for fixed $\upsilon$, one has
\begin{equation}
\zeta'(x_i)=-2\pi+O\left(e^{-2\pi\varrho}\right).\label{eq:sec6-zeta-prime-edge}
\end{equation}
Hence the candidate mixed generator takes the universal edge form
\begin{equation}
\mathcal D^{\mathrm{cand}}_{A,\mathrm{mix}}=\partial_\varrho+2\pi\sum_{i=1}^{2}\bigl(\frac12+p_i^+\partial_{p_i^+}\bigr)+O\left(e^{-2\pi\varrho}\right).\label{eq:sec6-edge-generator}
\end{equation}
This has several immediate consequences. First, fixed $p_i^+$ sectors are not closed under the candidate modular flow.  Even if the endpoint variables translate rigidly in $\varrho$, the mixed representation necessarily remembers the rescaling of the null directions through the operators $p_i^+\partial_{p_i^+}$. Second, a field at fixed total $P^+=p_1^++p_2^+$ is not, in general, modularly closed.  The modular flow naturally acts on the full $p_i^+$-space, not on a fixed $P^+$ slice.  Thus any exact modular statement for the mixed representation should really be formulated before freezing $P^+$, or in a basis adapted to the dilations. Third, the natural basis adapted to \eqref{eq:sec6-edge-generator} is not the $p_i^+$ basis but the Mellin basis in $p_i^+$, because $p_i^+\partial_{p_i^+}$ acts diagonally there. Accordingly, define the Mellin-transformed edge bilocal
\begin{equation}
\widehat\eta^{\mathrm{edge}}_{\sigma_1,\sigma_2}(x^+;\varrho,\upsilon)=\int_0^\infty \frac{dp_1^+}{p_1^+}\frac{dp_2^+}{p_2^+}\,(p_1^+)^{-i\sigma_1}(p_2^+)^{-i\sigma_2}\,\widehat\eta^{\mathrm{edge}}(x^+,p_1^+,p_2^+;\varrho,\upsilon).\label{eq:sec6-mellin-definition}
\end{equation}
On this Mellin sector, the edge generator \eqref{eq:sec6-edge-generator} becomes
\begin{equation}
\mathcal D^{\mathrm{cand}}_{A,\mathrm{Mellin}}=\partial_\varrho+2\pi\bigl(i\sigma_1+i\sigma_2+1\bigr)+O\left(e^{-2\pi\varrho}\right).\label{eq:sec6-edge-generator-mellin}
\end{equation}
This shows how to isolate the pure translation piece.  Define the renormalized Mellin edge operator
\begin{equation}
\widehat\eta^{\mathrm{edge,red}}_{\sigma_1,\sigma_2}(x^+;\varrho,\upsilon)=
e^{-2\pi(1+i\sigma_1+i\sigma_2)\varrho}\,
\widehat\eta^{\mathrm{edge}}_{\sigma_1,\sigma_2}(x^+;\varrho,\upsilon).\label{eq:sec6-edge-renormalized}
\end{equation}
If the true modular flow agrees with the candidate edge lift at leading order, then $\widehat\eta^{\mathrm{edge,red}}_{\sigma_1,\sigma_2}$ will transform by pure translation in $\varrho$, up to corrections suppressed by $e^{-2\pi\varrho}$.

It is worth emphasizing the conceptual point:  the reduced endpoint analysis by itself gives exact translation in $\varrho$.  The mixed representation then forces one to account for momentum dilations.  The Mellin transform is precisely the basis that diagonalizes those dilations.  Once this is done, the large-$\varrho$ problem again takes the form of approximate translation in $\varrho$.

\subsection{Fixed-spin sectors and bulk reconstruction}\label{subsec:fixed-spin-logical-status}

We can now summarize what has been established. The exact subregion statement for the reduced master field is \eqref{eq:sec6-fixed-theta-master-in-algebra}--\eqref{eq:sec6-fixed-theta-equality}.  This statement is rigorous and follows directly from the endpoint map and the equality proved in Section~\ref{HKLLequivalence}.  The exact reduced endpoint organization is furnished by \eqref{eq:auto:0030}--\eqref{eq:upsilon-theta-exact}.  In particular, the variables $(\varrho,\upsilon)$ linearize the canonical endpoint-preserving flow, and the relation \eqref{eq:upsilon-theta-exact} identifies $\upsilon$ with the auxiliary angular variable $\theta$ on each monotonic branch.

The first nontrivial new feature enters when passing to the mixed representation.  Equations \eqref{eq:sec6-candidate-null-lift}--\eqref{eq:sec6-edge-generator} show that, on the natural lift of the reduced endpoint flow to the null plane, modular transport is necessarily accompanied by momentum dilations.  This has an important structural consequence for projected bulk fields: fixed $p_i^+$ sectors, and hence in general fixed $P^+$ sectors and fixed $\theta$ sectors, are not expected to be exactly modularly closed.  The true modular organization of the mixed representation is therefore naturally expressed either before fixing $(P^+,\theta)$, or after Mellin transform in the lightfront momenta. This substantially limits the extent to which modular-flow considerations can directly illuminate the bilocal reconstruction formula, which is naturally organized at fixed $P^+$ and $\theta$.

Finally, what is the status of fixed-spin sectors?  The reconstructed master field is obtained from the mixed-representation bilocal by the map \eqref{eq:Phi-eta-map}, and the fixed-spin components are obtained by the angular projection \eqref{eq:projection-formula}.  However, in the subregion only a limited $\theta$-range is available at fixed $(X,Z)$ and the angular projection is no longer a purely local operation. Since different $p_i^+$ mix, fields with different $\theta$ mix and the separation of the fixed-spin sectors becomes a subtle issue.

To conclude this section, we have explained the sense in which bilocal holography yields an exact subregion reconstruction statement for the reduced master field, an exact reduced endpoint organization of the interval geometry, and a near-edge modular framework in the mixed representation.

\section{Conclusions}\label{conclusions}

In this paper we have developed a detailed formulation of bulk reconstruction within the framework of bilocal holography. The central result is an explicit map from the bilocal collective field to local bulk fields, obtained by expanding in single-trace primaries and resumming their descendants. The resulting reconstruction formula is remarkably local: the bulk field at a point is expressed in terms of bilocal data evaluated at a sharply defined configuration, with no residual non-local smearing over the boundary. This provides a concrete realization of how locality in the bulk can emerge from non-local, but gauge-invariant, degrees of freedom in the boundary theory.

A key achievement of this work is to place the bilocal reconstruction on firm footing through a series of stringent consistency checks. We have shown that the reconstructed field satisfies the correct bulk equations of motion and obeys the appropriate boundary conditions dictated by the GKPW prescription. By standard uniqueness arguments, this already implies equivalence with other reconstruction schemes once the same reduced variables, boundary data and Lorentzian prescription are used. We have gone further and verified this equivalence explicitly by comparing with the light-front HKLL construction. In the scalar sector the comparison uses the projected and descendant-resummed scalar block, not the bare local scalar primary. In the higher-spin sector the HKLL radial phase is retained until it is identified with the relative transverse phase of the separated bilocal endpoints. In this sense, bilocal holography provides not merely an alternative representation of bulk reconstruction, but a genuinely constructive one in which the mechanism of emergence of bulk fields can be followed step by step.

Another important outcome concerns the structure of the angular (or helicity) decomposition. The bilocal construction naturally organizes the bulk field in terms of angular harmonics, but we have emphasized that the relation between current spin and angular momentum is triangular: higher-spin current blocks contribute to lower angular harmonics. The explicit formulas derived in this paper isolate the diagonal, highest-harmonic contributions, and make clear how the full bulk field is assembled from these ingredients. 

We have also explored the implications of the reconstruction for subregion duality. Restricting the bilocal to a boundary subregion, we find that the reconstructable bulk region agrees precisely with the entanglement wedge, in accord with general expectations. This provides a concrete realization of entanglement wedge reconstruction within a fully explicit and computable framework. We further examined modular-flow-inspired constructions, highlighting both their promise and their limitations in the mixed lightfront representation. In particular, the necessity of momentum dilations in the lifted flow indicates that fixed lightfront momentum sectors are not modularly closed, suggesting that the natural language for modular organization involves either unfixed momentum variables or their Mellin transform.

There are several directions in which the present work can be extended. A natural next step is to develop a fully diagonal basis for the angular decomposition, in which the triangular mixing between different current blocks is removed by an explicit field redefinition. It would also be interesting to understand the extent to which the locality observed here persists beyond the large-$N$ or free-field limits, and how interactions deform the reconstruction formula.

Another important direction is to deepen the connection with modular flow. While exact modular Hamiltonians are difficult to construct in general, the bilocal framework may provide a tractable arena in which approximate or effective modular flows can be studied systematically. This could shed light on the dynamical origin of entanglement wedges and on the interplay between kinematics and dynamics in subregion duality. In this context, the study of bulk reconstruction for disconnected boundary subregions is important. 

Finally, the constructive nature of bilocal holography suggests broader applications. In particular, it offers a promising framework for addressing questions about the fine-grained structure of bulk locality, the emergence of causal structure, and the organization of degrees of freedom in quantum gravity. More speculatively, the explicit control over boundary variables may make it possible to probe regimes—such as finite $N$ or strongly coupled sectors—where other approaches to bulk reconstruction are less effective. Taken together, these results indicate that bilocal holography is not only a consistent realization of bulk reconstruction, but also a powerful tool for uncovering the detailed mechanisms by which spacetime geometry emerges from quantum degrees of freedom.

\begin{center} 
{\bf Acknowledgements}
\end{center}
RdMK would like to acknowledge very helpful discussions with Antal Jevicki.  The work of RdMK, AG and AR was supported by a start up research fund of Huzhou Normal University, a Zhejiang Province talent award and by a Changjiang Scholar award. M.K. was supported by supported by the NRF grants 2021R1A2C2012350 and RS-2025-25414114.

\appendix

\section{The standard Poisson-kernel construction}\label{poissonkernel}

The standard Poisson-kernel construction is the canonical way to solve a boundary-value problem in a half-space by writing the bulk field as a boundary convolution. The cleanest example is the Laplace equation in the upper half-space
\begin{equation}
\mathbb R^d_+ = \{(x,z)\in \mathbb R^{d-1}\times \mathbb R_{>0}\},\qquad\Delta u = (\partial_z^2+\Delta_x)u = 0,
\label{eq:auto:0039}
\end{equation}
with boundary data $u(x,0)=f(x)$. One looks for a kernel $P_z(x-x')$ such that
\begin{equation}
u(x,z)=\int_{\mathbb R^{d-1}} P_z(x-x')\,f(x')\,dx'
\label{eq:auto:0040}
\end{equation}
is harmonic, approaches $f$ as $z\to 0^+$, and is regular for $z>0$. The standard derivation is by Fourier transform in the boundary coordinates. Write
\begin{equation}
u(x,z)=\int \frac{d^{d-1}k}{(2\pi)^{d-1}}\,e^{ik\cdot x}\,\widehat u(k,z).
\label{eq:auto:0041}
\end{equation}
Then the Laplace equation becomes an ODE in $z$
\begin{equation}
(\partial_z^2-|k|^2)\widehat u(k,z)=0.
\label{eq:auto:0042}
\end{equation}
Its general solution is
\begin{equation}
\widehat u(k,z)=A(k)e^{|k|z}+B(k)e^{-|k|z}.
\label{eq:auto:0043}
\end{equation}
For a regular solution in the upper half-space, discard the growing mode to find
\begin{equation}
\widehat u(k,z)=\widehat f(k)e^{-|k|z},
\label{eq:auto:0044}
\end{equation}
because the boundary condition at $z=0$ requires $\widehat u(k,0)=\widehat f(k)$. The solution is
\begin{equation}
u(x,z)=\int \frac{d^{d-1}k}{(2\pi)^{d-1}}\,e^{ik\cdot x}e^{-|k|z}\widehat f(k).
\label{eq:auto:0045}
\end{equation}
This is already the Poisson-kernel representation in Fourier space. The Poisson kernel itself is the inverse Fourier transform of $e^{-|k|z}$
\begin{equation}
P_z(x)=\int \frac{d^{d-1}k}{(2\pi)^{d-1}}\,e^{ik\cdot x}e^{-|k|z}.
\label{eq:auto:0046}
\end{equation}
Evaluating that transform gives the familiar closed form
\begin{equation}
P_z(x)=c_d\,\frac{z}{(|x|^2+z^2)^{d/2}},
\label{eq:auto:0047}
\end{equation}
where $c_d$ is chosen so that
\begin{equation}
\int_{\mathbb R^{d-1}} P_z(x)\,dx = 1.
\label{eq:auto:0048}
\end{equation}
Thus
\begin{equation}
u(x,z)=c_d\int_{\mathbb R^{d-1}}\frac{z}{(|x-x'|^2+z^2)^{d/2}}\,f(x')\,dx'.
\label{eq:auto:0049}
\end{equation}
This is the standard Poisson-kernel construction. In one boundary dimension, the kernel is especially simple:
\begin{equation}
P_z(x-x')=\frac{1}{\pi}\frac{z}{(x-x')^2+z^2},\label{eq:auto:0050}
\end{equation}
so
\begin{equation}
u(x,z)=\frac{1}{\pi}\int_{-\infty}^{\infty}\frac{z}{(x-x')^2+z^2}f(x')\,dx'.
\label{eq:auto:0051}
\end{equation}
This is the kernel that appears when we discuss the factor $e^{-Z|k|}$: its inverse Fourier transform is the one-dimensional Poisson kernel. A mixed-space solution of the form
\begin{equation}
\widehat u(k,z)=e^{-|k|z}\widehat f(k)
\label{eq:auto:0052}
\end{equation}
corresponds in position space to a nonlocal boundary smearing over $x'$.

\section{Chebyshev identity}\label{Chebyshevidentity}

In this Appendix we prove the identity
\begin{equation}
\pi\sum_{k=0}^{2s}\frac{(-1)^kA^{2s-k}B^k}{k!(2s-k)!\,\Gamma\left(k+\frac12\right)\Gamma\left(2s-k+\frac12\right)}
=\frac{4^{2s}}{(4s)!}(A+B)^{2s}T_{2s}\!\left(\frac{A-B}{A+B}\right)\label{eq:app-chebyshev-main}
\end{equation}
for commuting $A$ and $B$. Since both sides are polynomials in $A$ and $B$, it is enough to prove the identity for positive real $A$ and $B$; the polynomial identity then extends it to arbitrary commuting variables. Using
\begin{equation}
\frac{\pi}{k!(2s-k)!\,\Gamma\left(k+\frac12\right)\Gamma\left(2s-k+\frac12\right)}=\frac{4^{2s}}{(2k)!(4s-2k)!}
\label{eq:auto:0054}
\end{equation}
the left-hand side of \eqref{eq:app-chebyshev-main} becomes
\begin{equation}
\frac{4^{2s}}{(4s)!}\sum_{k=0}^{2s}(-1)^k\binom{4s}{2k}A^{2s-k}B^k.\label{eq:app-chebyshev-binomial}
\end{equation}
Assume temporarily that $A=\alpha^2$ and $B=\beta^2$ with $\alpha,\beta>0$. Then
\begin{equation}
(\alpha+i\beta)^{4s}+(\alpha-i\beta)^{4s}=2\sum_{k=0}^{2s}(-1)^k\binom{4s}{2k}\alpha^{4s-2k}\beta^{2k}
=2\sum_{k=0}^{2s}(-1)^k\binom{4s}{2k}A^{2s-k}B^k.
\label{eq:auto:0055}
\end{equation}
Thus
\begin{equation}
\sum_{k=0}^{2s}(-1)^k\binom{4s}{2k}A^{2s-k}B^k=\frac{(\alpha+i\beta)^{4s}+(\alpha-i\beta)^{4s}}{2}.
\label{eq:auto:0056}
\end{equation}
Write
\begin{equation}
\alpha+i\beta=\sqrt{A+B}\,e^{i\vartheta/2},\qquad\cos\vartheta=\frac{A-B}{A+B}.
\label{eq:auto:0057}
\end{equation}
Then
\begin{equation}
\frac{(\alpha+i\beta)^{4s}+(\alpha-i\beta)^{4s}}{2}=(A+B)^{2s}\cos(2s\vartheta)
=(A+B)^{2s}T_{2s}\!\left(\frac{A-B}{A+B}\right).
\label{eq:auto:0058}
\end{equation}
Substituting this into \eqref{eq:app-chebyshev-binomial} proves \eqref{eq:app-chebyshev-main}. Since both sides are polynomials in $A$ and $B$, the identity holds for arbitrary commuting $A$ and $B$.

\section{Proof of the coefficient identity}\label{identity}

In this Appendix we prove the coefficient formula \eqref{eq:sec3-hkll-coeff-explicit}. To do this we reuse the current algebra and descendant resummation already established in Section~\ref{etaprimary}, together with the mixed HKLL kernel derived in Section~\ref{HKLLequivalence}. The important point is that the radial HKLL factor is never discarded. We keep it explicitly, identify it with the bilocal endpoint phase on the two-particle support of the free vector model, and only then apply the descendant resummation.

\paragraph{The current generating polynomial:} Introduce the generating polynomial for the spin-$2s$ current,
\begin{equation}
\mathcal J_{2s}(x^+,x^-,x;y^-,y)\equiv\sum_{m=0}^{2s}\binom{2s}{m}(y^-)^m y^{2s-m}j_{2s}^{(m)}(x^+,x^-,x).
\label{eq:app-current-generating}
\end{equation}
For free scalars, the standard equal-$x^+$ current formula is
\begin{equation}
\mathcal J_{2s}(x^+,x^-,x;y^-,y)=\pi\sum_{a=1}^{N}\sum_{\ell=0}^{2s}
\frac{(-1)^\ell}{\ell!(2s-\ell)!\Gamma\!\left(\ell+\frac12\right)\Gamma\!\left(2s-\ell+\frac12\right)}
:\mathcal A^{\,2s-\ell}\phi^a\,\mathcal B^{\,\ell}\phi^a: ,\label{eq:app-free-current}
\end{equation}
where
\begin{equation}
\mathcal A\equiv y^-\partial_-^{(1)}+y\partial_{x_1},\qquad\mathcal B\equiv y^-\partial_-^{(2)}+y\partial_{x_2}.
\label{eq:app-A-B-raw}
\end{equation}
It is important that \eqref{eq:app-free-current} is a polynomial identity. We will use it only on mixed Fourier modes, where no ambiguity about operator ordering or phases remains. Moving to Fourier space we have
%
%Now consider a single mixed Fourier mode with momenta $(p_1^+,p_1)$ and $(p_2^+,p_2)$:
%\begin{equation}
%\phi(x_1^-,x_1)\phi(x_2^-,x_2)\;\propto\;
%e^{-ip_1^+x_1^- -ip_1x_1}e^{-ip_2^+x_2^- -ip_2x_2}.
%\label{eq:auto:0059}
%\end{equation}
%On such a mode,
\begin{equation}
\mathcal A\mapsto -i\,(p_1^+y^-+p_1y),\qquad\mathcal B\mapsto -i\,(p_2^+y^-+p_2y).
\label{eq:app-A-B-mode}
\end{equation}
Hence \eqref{eq:app-free-current} reduces modewise to the polynomial
\begin{align}
\mathcal J_{2s}^{\rm mode}(y^-,y)
&=
\pi\sum_{\ell=0}^{2s}
\frac{(-1)^\ell}{\ell!(2s-\ell)!\Gamma\!\left(\ell+\frac12\right)\Gamma\!\left(2s-\ell+\frac12\right)}
\nonumber\\
&\qquad\times
\bigl[-i(p_1^+y^-+p_1y)\bigr]^{2s-\ell}
\bigl[-i(p_2^+y^-+p_2y)\bigr]^\ell .
\label{eq:app-current-mode-poly}
\end{align}

\paragraph{Chebyshev form and light-front variables:} Introduce
\begin{equation}
P^+=p_1^++p_2^+,\qquad\nu\equiv\frac{p_1^+-p_2^+}{P^+}=\cos\theta,\qquad
\nu_\perp\equiv\sqrt{1-\nu^2}=\sin\theta,\label{eq:app-plus-defs}
\end{equation}
together with
\begin{equation}
k=p_1+p_2,\qquad q=\sqrt{\frac{p_2^+}{p_1^+}}\,p_1-\sqrt{\frac{p_1^+}{p_2^+}}\,p_2 .
\label{eq:app-k-q-def}
\end{equation}
Then
\begin{equation}
p_1=\frac{1}{2}\bigl[(1+\nu)k+\nu_\perp q\bigr],\qquad p_2=\frac{1}{2}\bigl[(1-\nu)k-\nu_\perp q\bigr].
\label{eq:app-p1-p2-inverse}
\end{equation}
Define the commuting mode polynomials
\begin{equation}
\alpha_k\equiv -i\,(P^+y^-+ky),\qquad\beta_{k,q}\equiv -i\bigl(P^+\nu\,y^-+(k\nu+q\nu_\perp)y\bigr).
\label{eq:app-AkBk}
\end{equation}
A direct substitution of \eqref{eq:app-p1-p2-inverse} gives
\begin{equation}
\frac{\alpha_k+\beta_{k,q}}{2}=-i\,(p_1^+y^-+p_1y),\qquad\frac{\alpha_k-\beta_{k,q}}{2}=-i\,(p_2^+y^-+p_2y).\label{eq:app-UV-eval}
\end{equation}
Using the polynomial identity proved in Appendix~\ref{Chebyshevidentity},
\begin{equation}
\pi\sum_{\ell=0}^{2s}\frac{(-1)^\ell A^{2s-\ell}B^\ell}{\ell!(2s-\ell)!\Gamma\!\left(\ell+\frac12\right)\Gamma\!\left(2s-\ell+\frac12\right)}=\frac{4^{2s}}{(4s)!}(A+B)^{2s}T_{2s}\!\left(\frac{A-B}{A+B}\right),\label{eq:app-chebyshev-again}
\end{equation}
with $A=\frac{\alpha_k+\beta_{k,q}}{2}$ and $B=\frac{\alpha_k-\beta_{k,q}}{2}$, we obtain
\begin{equation}
\mathcal J_{2s}^{\rm mode}(y^-,y)=\frac{4^{2s}}{(4s)!}\,\alpha_k^{\,2s}\,\,
T_{2s}\!\left(\frac{\beta_{k,q}}{\alpha_k}\right).\label{eq:app-current-mode-cheb}
\end{equation}
Thus the exact mode polynomial of the spin-$2s$ current is precisely the Chebyshev polynomial built from \eqref{eq:app-AkBk}.

\paragraph{Fourier representation of the HKLL coefficients:} The object on which the Chebyshev polynomial acts is the point-split bilinear of two free scalars,
\begin{equation}
\Sigma\bigl(x^+;x_1^-,x_1;x_2^-,x_2\bigr)=\sum_{a=1}^{N}:\phi^a(x^+,x_1^-,x_1)\,\phi^a(x^+,x_2^-,x_2):\,.
\label{eq:app-Sigma-pointsplit}
\end{equation}
Its full mixed Fourier transform is
\begin{eqnarray}
\widehat\Sigma(\omega;p_1^+,p_2^+;p_1,p_2)&=&\int dx^+\,dx_1^-\,dx_2^-\,dx_1\,dx_2\; e^{i\omega x^+}
 e^{ip_1^+x_1^-+ip_2^+x_2^-} e^{ip_1x_1+ip_2x_2}\cr\cr
 &&\qquad\times \Sigma\bigl(x^+;x_1^-,x_1;x_2^-,x_2\bigr).
\label{eq:app-Sigma-full-fourier}
\end{eqnarray}
The differential operators $\mathcal A$ and $\mathcal B$ act on the first and second scalar factors separately, so in Fourier space the exact current polynomial multiplies \eqref{eq:app-Sigma-full-fourier}. Here we work at fixed
\begin{equation}
P^+=p_1^++p_2^+,\qquad\nu=\frac{p_1^+-p_2^+}{P^+}=\cos\theta,\qquad k=p_1+p_2,
\label{eq:app-Sigma-fixed-kinematics}
\end{equation}
and then impose the HKLL radial relation $q=q(\omega,k)$. Accordingly we abbreviate
\begin{equation}
\widehat\Sigma(\omega,P^+,k)=\widehat\Sigma\bigl(\omega;p_1^+,p_2^+;p_1,p_2\bigr)\Big|_{\nu=\cos\theta,\;q=q(\omega,k)}.\label{eq:app-Sigma-abbrev}
\end{equation}
This is merely a shorthand; the suppressed dependence on the momentum split between the two scalar factors is encoded by $\theta$ and $q(\omega,k)$. With this notation, define
\begin{equation}
\Psi_Z(X^+,P^+,X)\equiv\int\frac{d\omega\,dk}{(2\pi)^2}\,e^{-i\omega X^+-ikX}\,e^{-iZq(\omega,k)}\,
\widehat\Sigma(\omega,P^+,k).\label{eq:app-psi-generic}
\end{equation}
The coefficient of $(y^-)^m y^{2s-m}$ in the fixed-spin HKLL master field is then
\begin{equation}
\Xi_{s,m}^{\rm HKLL}(X^+,P^+,X,Z;\theta)=\frac{1}{\binom{2s}{m}}[(y^-)^m y^{2s-m}]\left[\frac{4^{2s}}{(4s)!}
\,\alpha^{\,2s}T_{2s}\!\left(\frac{\beta}{\alpha}\right)\Psi_Z\right],\label{eq:app-Xi-def}
\end{equation}
where $\alpha$ and $\beta$ are the differential operators obtained from \eqref{eq:app-AkBk} by the replacements $k\mapsto i\partial_X$ and $q\mapsto i\partial_Z$. We now evaluate \eqref{eq:app-Xi-def} explicitly. Since
\begin{equation}
i\partial_X\,e^{-ikX}=k\,e^{-ikX},\qquad i\partial_Z\,e^{-iZq}=q\,e^{-iZq},\label{eq:app-fourier-action}
\end{equation}
the differential operators $\alpha$ and $\beta$ act on the Fourier kernel in \eqref{eq:app-psi-generic} by multiplication with the commuting mode polynomials $\alpha_k$ and $\beta_{k,q}$
\begin{eqnarray}
\alpha\Bigl(e^{-i\omega X^+-ikX}e^{-iZq}\Bigr)&=&\alpha_k\,e^{-i\omega X^+-ikX}e^{-iZq},\cr\cr
\beta\Bigl(e^{-i\omega X^+-ikX}e^{-iZq}\Bigr)&=&\beta_{k,q}\,e^{-i\omega X^+-ikX}e^{-iZq}.\label{eq:app-AB-on-kernel}
\end{eqnarray}
Therefore \eqref{eq:app-Xi-def} becomes
\begin{align}
\Xi_{s,m}^{\rm HKLL}&=\int\frac{d\omega\,dk}{(2\pi)^2}\,e^{-i\omega X^+-ikX}\,
\widehat\Xi_{s,m}^{\rm HKLL}(\omega,P^+,k,Z;\theta),\label{eq:app-Xi-fourier}
\end{align}
with Fourier-space integrand
\begin{align}
\widehat\Xi_{s,m}^{\rm HKLL}(\omega,P^+,k,Z;\theta)&=e^{-iZq(\omega,k)}\frac{1}{\binom{2s}{m}}[(y^-)^m y^{2s-m}]\left[
\frac{4^{2s}}{(4s)!}\,\alpha_k^{\,2s}\,\,T_{2s}\!\left(\frac{\beta_{k,q(\omega,k)}}{\alpha_k}\right)
\widehat\Sigma(\omega,P^+,k)\right].\label{eq:app-Xi-integrand}
\end{align}
Using \eqref{eq:app-current-mode-cheb}, this may be rewritten as
\begin{equation}
\widehat\Xi_{s,m}^{\rm HKLL}(\omega,P^+,k,Z;\theta)=e^{-iZq(\omega,k)}\frac{1}{\binom{2s}{m}}[(y^-)^m y^{2s-m}]
\Bigl[\mathcal J_{2s}^{\rm mode}(y^-,y)\,\widehat\Sigma(\omega,P^+,k)\Bigr].\label{eq:app-Xi-integrand-current}
\end{equation}
The phase in \eqref{eq:app-Xi-integrand-current} is essential, so let us first identify it with the bilocal endpoint phase. For a two-particle mode of the free scalar theory,
\begin{equation}
\omega=\frac{p_1^2}{2p_1^+}+\frac{p_2^2}{2p_2^+},\qquad
P^+=p_1^++p_2^+,
\qquad k=p_1+p_2 .
\label{eq:app-two-particle-energy}
\end{equation}
Using the relative momentum defined in \eqref{eq:app-k-q-def}, one finds
\begin{align}
2P^+\omega-k^2
&=P^+\left(\frac{p_1^2}{p_1^+}+\frac{p_2^2}{p_2^+}\right)-(p_1+p_2)^2\nonumber\\
&=\frac{p_2^+}{p_1^+}p_1^2+
\frac{p_1^+}{p_2^+}p_2^2-2p_1p_2
=\left(\sqrt{\frac{p_2^+}{p_1^+}}p_1-
\sqrt{\frac{p_1^+}{p_2^+}}p_2\right)^2
=q^2 .
\label{eq:app-qrel-onshell}
\end{align}
Thus, on the two-particle support of the bilocal field, the HKLL radial momentum is the relative transverse momentum,
\begin{equation}
q(\omega,k)=q,
\label{eq:app-radial-relative-identification}
\end{equation}
with the sign fixed by the radial branch selected by the Lorentzian prescription. Next, the bilocal-to-bulk map gives
\begin{equation}
x_1=X+Z\tan{\theta\over2},\qquad
x_2=X-Z\cot{\theta\over2}.
\label{eq:app-endpoint-map-phase}
\end{equation}
Since
\begin{equation}
\tan{\theta\over2}=\sqrt{\frac{p_2^+}{p_1^+}},
\qquad
\cot{\theta\over2}=\sqrt{\frac{p_1^+}{p_2^+}},
\label{eq:app-half-angle-plus}
\end{equation}
we obtain
\begin{equation}
p_1x_1+p_2x_2=kX+qZ.
\label{eq:app-bilocal-phase-identity}
\end{equation}
Consequently
\begin{equation}
e^{-ikX}e^{-iZq(\omega,k)}=e^{-i(p_1x_1+p_2x_2)}.
\label{eq:app-HKLL-phase-as-bilocal-phase}
\end{equation}
After imposing the two-particle support, the HKLL radial phase is precisely the phase which translates the coincident current data into the separated bilocal endpoints. With this identification, \eqref{eq:app-Xi-integrand-current} is the same mixed Fourier-space fixed-$m$ contribution that appears in the descendant resummation of Section~\ref{etaprimary}. Translating the result to the present bulk variables,
\begin{equation}
\nu=\cos\theta,\qquad Z=y\sin\theta,\qquad X=x+\nu y,\qquad\partial_x=\partial_X,
\label{eq:app-bulk-variable-map}
\end{equation}
one finds
\begin{align}
\widehat\Xi_{s,m}^{\rm HKLL}(\omega,P^+,k,Z;\theta)&=2\pi\,2^{3-2s}(2s)!\sum_{r=0}^{m}\binom{m}{r}(2i)^r r!\,
\frac{Z^{2s-r}}{(P^+)^{m}}(\sin\theta)^{2s-r}\nonumber\\[1mm]
&\qquad\times P_r^{\left(2s-r-\frac12,\;2s-r-\frac12\right)}(\cos\theta)\,(-1)^{m-r}k^{m-r}\,
\widehat j_{2s}^{(m)}(\omega,P^+,k).\label{eq:app-Xi-kspace}
\end{align}
Substituting \eqref{eq:app-Xi-kspace} into \eqref{eq:app-Xi-fourier} gives
\begin{align}
\Xi_{s,m}^{\rm HKLL}&=2\pi\,2^{3-2s}(2s)!\sum_{r=0}^{m}\binom{m}{r}(2i)^r r!\,\frac{Z^{2s-r}}{(P^+)^{m}}
(\sin\theta)^{2s-r}P_r^{\left(2s-r-\frac12,\;2s-r-\frac12\right)}(\cos\theta)\nonumber\\[1mm]
&\qquad\times \int\frac{d\omega\,dk}{(2\pi)^2}\,
e^{-i\omega X^+-ikX}(-1)^{m-r}k^{m-r}\widehat j_{2s}^{(m)}(\omega,P^+,k).\label{eq:app-Xi-before-integrals}
\end{align}
We now perform the inverse Fourier transforms explicitly. By definition,
\begin{equation}
j_{2s}^{(m)}(X^+,P^+,X)=\int\frac{d\omega\,dk}{(2\pi)^2}\,e^{-i\omega X^+-ikX}\,
\widehat j_{2s}^{(m)}(\omega,P^+,k),\label{eq:app-current-inverse-fourier}
\end{equation}
so the $\omega$-integral in \eqref{eq:app-Xi-before-integrals} first reconstructs the mixed representation,
\begin{equation}
\int\frac{d\omega}{2\pi}\,e^{-i\omega X^+}\widehat j_{2s}^{(m)}(\omega,P^+,k)=j_{2s}^{(m)}(X^+,P^+,k),
\label{eq:app-omega-integral}
\end{equation}
and then the $k$-integral is evaluated using
\begin{equation}
\partial_X^{n}e^{-ikX}=(-ik)^n e^{-ikX}\qquad\Longrightarrow\qquad\int\frac{dk}{2\pi}\,e^{-ikX}k^n\widehat f(k)
= i^n\partial_X^n f(X).\label{eq:app-k-integral-identity}
\end{equation}
Hence
\begin{align}
\int\frac{d\omega\,dk}{(2\pi)^2}\,e^{-i\omega X^+-ikX}(-1)^{m-r}k^{m-r}\widehat j_{2s}^{(m)}(\omega,P^+,k)
&=(-1)^{m-r}i^{m-r}\partial_X^{m-r}j_{2s}^{(m)}(X^+,P^+,X)\nonumber\\
&=(-i)^{m-r}\partial_X^{m-r}j_{2s}^{(m)}(X^+,P^+,X).\label{eq:app-omega-k-evaluated}
\end{align}
Inserting \eqref{eq:app-omega-k-evaluated} into \eqref{eq:app-Xi-before-integrals}, we obtain
\begin{align}
\Xi_{s,m}^{\rm HKLL}&=2\pi\,2^{3-2s}(2s)!\sum_{r=0}^{m}\binom{m}{r}(2i)^r(-i)^{m-r}r!\,\frac{Z^{2s-r}}{(P^+)^{m}}
(\sin\theta)^{2s-r}\nonumber\\[1mm]
&\qquad\qquad\times P_r^{\left(2s-r-\frac12,\;2s-r-\frac12\right)}(\cos\theta)\,
\partial_X^{m-r}j_{2s}^{(m)}(X^+,P^+,X).\label{eq:app-Xi-final}
\end{align}
This is exactly equation \eqref{eq:sec3-hkll-coeff-explicit}, which is what we wanted to prove. The coefficient formula follows by rewriting the HKLL integrand through the exact free-field current mode polynomial and then applying the descendant resummation already carried out in Section~\ref{etaprimary}; the final additional step is the explicit evaluation of the $\omega$- and $k$-integrals displayed above.

\section{Scalar sector: local operator vs. scalar conformal block}\label{scalarsupport}

It is important to sharply distinguish between two different objects which
might naively be conflated
\begin{itemize}
\item[(i)] the local scalar primary
\begin{equation}
j_0^{(0)}(x)=:\phi^a(x)\phi^a(x):,
\end{equation}
\item[(ii)] the scalar ($s=0$) conformal block appearing in the bilocal
operator product expansion.
\end{itemize}
These are \emph{not} the same object, and in particular the local operator
$j_0^{(0)}$ does \emph{not} satisfy a free equation of motion. The purpose of the notation $\mathcal J_0^{\rm blk}$ in the main text is precisely to distinguish the scalar block datum from the bare local primary.

\paragraph{The local operator is not free.}
Starting from the free equation for each fundamental field,
\begin{equation}
2\partial_+\partial_-\phi^a=\partial_x^2\phi^a,
\end{equation}
we find
\begin{equation}
\left(2\partial_+\partial_- - \partial_x^2\right):\phi^a\phi^a:
=4:\partial_+\phi^a\,\partial_-\phi^a:-2:\partial_x\phi^a\,\partial_x\phi^a:,
\end{equation}
which is non-zero in general. Thus the local composite $j_0^{(0)}$ is \emph{not} a free field.

\paragraph{Scalar projection.} The relevant object in the bilocal formalism is not the local operator $j_0^{(0)}$ itself, but the scalar conformal block obtained from the equal-$x^+$ bilocal operator
\begin{equation}
\sum_a:\phi^a(x+y)\phi^a(x-y):.
\end{equation}
Expanding in the relative coordinate $y^\mu$, one has standard OPE
decomposition
\begin{equation}
\sum_a:\phi^a(x+y)\phi^a(x-y):= \sum_{s=0}^\infty\sum_{d=0}^\infty
c_{sd}\,(y^\mu\partial_\mu)^{2d}y^{\mu_1}\cdots y^{\mu_{2s}}
j^{(2s)}_{\mu_1\cdots\mu_{2s}}(x).
\end{equation}
The \emph{scalar projection} is the extraction of the $s=0$ conformal
block:
\begin{equation}
\eta_{(0)}=\sum_{d=0}^\infty c_{0d}\,(y^\mu\partial_\mu)^{2d}j_0^{(0)}(x).
\end{equation}
This is a highly nontrivial object: it contains the full descendant tower of the scalar primary.

\paragraph{Descendant resummation.} Upon Fourier transforming in the relative light-front coordinate $y^-$, the descendant tower resums to a closed expression,
\begin{equation}
\eta_{(0)}(x^+,P^+,q^+;x,y)=\frac{4}{P^+\sqrt{1-\nu^2}}\,
j_0^{(0)}(x^+,P^+;x+\nu y),
\end{equation}
where $\nu=\cos\theta$ as usual. Thus, \emph{after} resummation, the scalar block is represented by a rigid translation of the local operator. We denote the corresponding reduced scalar-block datum, after projection and descendant resummation, by $\mathcal J_0^{\rm blk}$.

\paragraph{Bilocal-to-bulk map.}
Applying the bilocal-to-bulk change of variables
\begin{equation}
\nu=\cos\theta,\qquad y=\frac{Z}{\sin\theta},\qquad x+\nu y=X,
\end{equation}
one obtains
\begin{equation}
\Phi^{\rm biloc}\big|_{\eta^{(0)}}=8\pi\mathcal J_0^{\rm blk}(X^+,P^+,X),
\end{equation}
which is manifestly independent of $Z$.

\paragraph{Key conceptual point.}
The $Z$-independence and hence the effective free center-of-mass dynamics emerges \emph{only after}
\begin{enumerate}
\item projecting onto the $s=0$ conformal block, and
\item resumming the full tower of descendants.
\end{enumerate}
It is incorrect to attribute this free field property to the local operator $j_0^{(0)}$ itself. The simplification is a property of the \emph{scalar conformal block of the bilocal field}, not of the bare composite operator. Equivalently, the free equation used in the HKLL comparison is an equation for $\mathcal J_0^{\rm blk}$, not for $j_0^{(0)}$.

\paragraph{Consequence.}
Because the resulting contribution to the master field is independent of $Z$, the blockwise bulk equation reduces to the center-of-mass light-front equation
\begin{equation}
\left(i\partial_{X^+}+{1\over 2P^+}\partial_X^2\right)\mathcal J_0^{\rm blk}(X^+,P^+,X)=0 .
\end{equation}
Here ``blockwise'' means that we first project the bilocal onto the scalar conformal module and only then use the linear bilocal equation of motion. This is legitimate because the conformal-block decomposition is into independent modules and the free light-front evolution preserves this decomposition. Fourier transforming gives
\begin{equation}
\left(2P^+\omega-k^2\right)\widehat {\mathcal J}_0^{\rm blk}(\omega,P^+,k)=0,
\end{equation}
which implies that the HKLL radial momentum vanishes on the support of the scalar block. This statement should thus be understood as a property of the projected and resummed bilocal scalar sector, rather than of the local primary operator.

\end{document}